\documentclass[nofootinbib,prd,twocolumn,showpacs,showkeys,preprintnumbers]{revtex4-1}
\usepackage{hyperref,amssymb,amsmath,mathrsfs,bm,graphicx}
\usepackage[dvipsnames]{xcolor}
\begin{document}

\title {Gravitational Cracking of General Relativistic Polytropes: a  generalized scheme}

\author{P. Le\'on}
\email{pablo.leon@ua.cl}
\affiliation{Departamento de F\'isica, Universidad de Antofagasta, Aptdo 02800, Chile.}

\author{E. Fuenmayor}
\email{ernesto.fuenmayor@ciens.ucv.ve}
\affiliation{Centro de F\'isica Te\'orica y Computacional,\\ Escuela de F\'isica, Facultad de Ciencias, Universidad Central de Venezuela, Caracas 1050, Venezuela}

\author{E. Contreras }
\email{econtreras@usfq.edu.ec}
\affiliation{Departamento de F\'isica, Colegio de Ciencias e Ingenier\'ia, Universidad San Francisco de Quito,  Quito 170901, Ecuador\\}

\begin{abstract}
We discuss the effect that small fluctuations of both local anisotropy and energy density, may have on the occurrence of cracking in spherical
compact objects satisfying a polytropic equation of state. A systematic scheme to bring the fluid configurations out of hydrostatic equilibrium is revisited. Various models of polytropes are considered and it is shown that departures from equilibrium may lead to the appearance of cracking  for a wide range of values of the parameters involved. Prospective applications of the obtained results to some astrophysical scenarios are pointed out.
\end{abstract}


\maketitle

\section{Introduction}
The concept of cracking was introduced by L. Herrera in \cite{cn} (and fine-tuned in later works \cite{cnbis, cn1, cn2, cn4, cn5, cn7bis, cn7bbis, mardam, 4p}), and corresponds to the situation in which radial forces of different signs appear within the system after it is perturbed.
We say that there is a cracking whenever, on a time smaller than the hydrostatic time scale, the radial force is directed inward in the inner part of the sphere and reverses its sign beyond some value of the radial coordinate. In the opposite case, we shall say that there is an overturning. As should be clear at this point, the concept of cracking is closely related to  the problem of structure formation \cite{cn3, cn6, cn7}. In \cite{cnbis} it was shown that cracking results only if the local anisotropy is perturbed suggesting that fluctuations of 
such a quantity
may be crucial in the occurrence of cracking. Even more, the appearance of cracking 
in initially isotropic configurations,
shows that even small deviations from local isotropy may lead to drastic changes in the evolution of the system \cite{cn1}. In any case, the relationship between anistropy and cracking seems doubtless.\\

It is well known that deviations of the isotropy and fluctuations of the local anisotropy in pressures may be caused by a large variety of physical phenomena of the kind  we expect to find in compact objects (see Refs. \cite{CHH,14, hmo02, 04, LH-C3, hod08, hsw08, p1, p2, anis1, anis2, anis3, anis4bis}, for an extensive discussion on this point. See also, \cite{o1,o2,o3,o4,o5} for recent developments). Among all these possibilities we would like to mention two which might be particularly related to our primary interest: i) intense magnetic field observed in compact objects such as white dwarfs, neutron stars, or magnetized strange quark stars  (see, for example, Refs. \cite{15, 16, 17, 18, 19,23, 24, 25, 26}) and ii) viscosity (see \cite{Anderson, sad, alford, blaschke, drago, jones, vandalen, Dong} and references therein). 
Besides, as  it  has  been  recently  proven, the isotropic pressure condition becomes unstable by the presence  of  dissipation,  energy  density  inhomogeneities and  shear \cite{LHP}. These points mentioned above can  explain  the  renewed  interest in the study of fluids not satisfying the isotropic condition and justify our present work which is based on anisotropic polytropes.\\

The general formalism to study polytropes for anisotropic matter was presented in both the Newtonian  \cite{1p} and the general relativistic regimes  \cite{2p, 3p} (for recent developments see \cite{5p,6p,7p,8lp,9lp} and
 for polytropes in other contexts see \cite{z1,z2,z3,z4}, for example) motivated by the fact that the polytropic equations of state allow to deal with a variety of fundamental astrophysical problems (see Refs. \cite{cha, sch,  2, 3, 7', 9, 8,10, 4a, 4b, 4c, 5, 6, 7, 11, 12, 13, pol1n}). The theory of polytropes is based on the polytropic equation of state, which can be written as one of the following possibilities
\begin{equation}P_r=K\rho_0^{\gamma}=K\rho_0^{1+1/n} 
\label{p1},
\end{equation} 
\begin{equation}
P_r=K\rho^{\gamma}=K\rho^{1+1/n} 
\label{p2},
\end{equation} 
where $P_{r}$, $\rho_0$ and $\rho$ denote the radial pressure, the  mass (baryonic) density and the energy density,  respectively. Constants $K$, $\gamma$, and $n$ are usually called  the polytropic constant, polytropic exponent, and polytropic index, respectively. 

The fact that the  principal stresses are unequal produces an extra indeterminacy so the introduction of an additional  condition to close the system is compulsory \cite{8p, 9p}. For example, in \cite{1p,2p,anis4}
it was considered a particular
ansatz which allowed to obtain an anisotropic model continually
linked with the isotropic case. Another interesting choice for the local pressure anisotropy was introduced in \cite{5p, 6p} where the main idea was the additional assumption that both principal stresses satisfy polytropic equations of state.\\

Other approaches to solve the field equations consist in to impose conditions on the metric variables. The vanishing of the Weyl tensor  (usually referred to as the conformally flat condition) is one of them \cite{Herrera2001}. Another one is the so called class one condition \cite{karmarkar} (for recent developments see, \cite{maurya, tello2,tello3,tello4,Nunez,tello5,tello6}, for example) which allows to construct solutions embedded in a $5$ dimensional Euclidean space (see \cite{7p} for details). Although both conformally flat and class one conditions have been broadly implemented to
close the system of Einstein's equations, the use of a new concept of complexity based on the scalar $Y_{TF}$ appearing in the orthogonal splitting of the Riemann tensor has increased in recent years \cite{VC}. The complexity factor in \cite{VC} is quite different from previous ones given in references \cite{VC1,VC2,VC3,VC4,VC5,VC6} in the sense that $Y_{TF}$ contains contributions from the energy density inhomogeneity and the local pressure anisotropy, combined in a very specific way, which vanishes for the homogeneous and locally isotropic fluid distribution which is considered the less complex system. It is worth mentioning that the complexity factor so defined not only vanishes for the simple configuration mentioned above but also may vanish when the two terms appearing in its definition cancel each other. So, vanishing complexity may correspond to very different systems \cite{VC, VC1}. It is worth mentioning that a very recent study establishes a relationship between families of solutions that have different complexities with the possible occurrence of cracking in the framework of gravitational decoupling \cite{GPRD}.\\ 

In this work, to study the appearance of cracking for our adopted polytropes models, we generalize the scheme proposed in \cite{cn4} in order to effectively break the hydrostatic equilibrium of the system. Doing so, it should be clear that assuming this lack of response of the fluid, i.e. the inability to adapt its radial pressure to the perturbed situation, is equivalent to assuming that the pressure-density relation (the ratio of specific heats) never reaches the value required for neutral equilibrium. Usually, the study of cracking depends on handpicking two parameters of the models that could be used to perform the energy density and local anisotropy perturbations. The advantage of our perturbation scheme, which is inspired on the developments in \cite{cn4}, is that it will be valid for any spherically symmetric internal solution of Einstein’s equations obeying a barotropic/polytropic equations of state and is independent of the particular model under study. 
It should be emphasised that, we shall restrict here to the case described by Eq. (\ref{p2}), for simplicity, just as it was developed and justified in reference \cite{6p}. In particular, we shall consider the conformally flat, the double, and the Karmarkar class I polytropes models reported in \cite{3p,6p,7p}. For completeness of this work, we also consider the recent vanishing complexity polytrope \cite{VC}.
 


This work is organized as follows. In the next section we study the basic equations of general relativity as well as a summary of the theory of relativistic polytropes. We dedicate Section 3 to introduce the general perturbation scheme raised in this work. In Section 4 we study the appearance of cracking for the polytrope models considered. Finally, Sections 5 and 6 are devoted to the results and concluding remarks, respectively.

\section{General Relativistic polytropes}

\subsection{Relevant equations and conventions}

Let us consider a static, spherically symmetric distribution of an  anisotropic fluid bounded by a surface $\Sigma$.  In Schwarzschild--like coordinates, the metric is given by
\begin{eqnarray}\label{metric}
ds^{2}=e^{\nu(r)}dt^{2}-e^{\lambda(r)}dr^{2}-r^{2}(
d\theta^{2}+\sin^{2}d\phi^{2}),
\end{eqnarray}
where $\nu$ and $\lambda$ are functions of $r$. 

The matter content of the sphere is described by the energy--momentum tensor 
\begin{eqnarray}
T_{\mu\nu}=(\rho+P_{\perp})u_{\mu}u_{\nu}-P_{\perp}g_{\mu\nu}+(P_{r}-P_{\perp})s_{\mu}s_{\nu},
\end{eqnarray}
where, 
\begin{eqnarray}
u^{\mu}=(e^{-\nu/2},0,0,0),
\end{eqnarray}
is the four velocity of the fluid, $s^{\mu}$ is defined as
\begin{eqnarray}
s^{\mu}=(0,e^{-\lambda},0,0),
\end{eqnarray}
with the properties $s^{\mu}u_{\mu}=0$, $s^{\mu}s_{\mu}=-1$ (we are assuming geometric units $c=G=1$). The metric (\ref{metric}), has to satisfy the Einstein field equations, which are given by
\begin{eqnarray}
\rho&=&-\frac{1}{8\pi}\bigg[-\frac{1}{r^{2}}+e^{-\lambda}\left(\frac{1}{r^{2}}-\frac{\lambda'}{r}\right) \bigg],\label{ee1}\\
P_{r}&=&-\frac{1}{8\pi}\bigg[\frac{1}{r^{2}}-e^{-\lambda}\left(
\frac{1}{r^{2}}+\frac{\nu'}{r}\right)\bigg],\label{ee2}
\end{eqnarray}
\begin{equation}
P_{\perp}=\frac{1}{8\pi}\bigg[ \frac{e^{-\lambda}}{4}
\left(2\nu'' +\nu'^{2}-\lambda'\nu'+2\frac{\nu'-\lambda'}{r}
\right)\bigg]\label{ee3},
\end{equation}
where primes denote derivative with respect to $r$.

Outside the fluid distribution the spacetime is given by the Schwarzschild exterior solution, namely
\begin{eqnarray}
ds^{2}&=&\left(1-\frac{2M}{r}\right)dt^{2}-\left(1-\frac{2M}{r}\right)^{-1}dr^{2}\nonumber\\
&&-r^{2}(d\theta^{2}+\sin^{2}d\phi^{2}).
\end{eqnarray}

Furthermore, we require the continuity of the first and  the second fundamental form across the boundary surface $r=r_{\Sigma}=\rm constant$, which implies,
\begin{eqnarray}
e^{\nu_{\Sigma}}&=&1-\frac{2M}{r_{\Sigma}},\label{nursig}\\
e^{-\lambda_{\Sigma}}&=&1-\frac{2M}{r_{\Sigma}}\label{lamrsig}\\
P_{r_{\Sigma}}&=&0,
\end{eqnarray}
where the subscript $\Sigma$ indicates that the quantity is evaluated at the boundary  surface $\Sigma$.

From the radial component of the conservation law,
\begin{eqnarray}\label{Dtmunu}
\nabla_{\mu}T^{\mu\nu}=0,
\end{eqnarray}
one obtains  the generalized Tolman--Oppenheimer--Volkoff equation for anisotropic matter  which reads,
\begin{eqnarray}\label{TOV}
\mathcal{R} \equiv P_{r}'+\frac{\nu'}{2}(\rho +P_{r})-\frac{2}{r}(P_{\perp}-P_{r})=0,
\end{eqnarray}
where $\mathcal{R}$ defines the total radial force on each fluid element. Alternatively, using
\begin{eqnarray}
\nu'=2\frac{m+4\pi P_{r}r^{3}}{r(r-2m)},
\end{eqnarray}
where the mass function $m$ is defined through
\begin{eqnarray}
e^{-\lambda}=1-2m/r\label{mf},
\end{eqnarray}
or, equivalently as
\begin{eqnarray}
m=4\pi \int_{0}^{r} \tilde{r}^2 \rho d\tilde{r},
\label{mf2}
\end{eqnarray}
we may rewrite Eq. (\ref{TOV}) in the form
\begin{eqnarray}\label{TOV2}
P_{r}'=-\frac{m+4\pi r^{3} P_{r}}{r(r-2m)}(\rho+P_{r})+\frac{2}{r}\Delta, \label{TOVb}
\end{eqnarray}
where
\begin{eqnarray}\label{delta}
\Delta=P_{\perp}-P_{r},
\end{eqnarray}
measures  the anisotropy of the system.

It is important to note that, on the one hand, if the fluid is in equilibrium the radial pressure gradient is balanced by a gravitational term that contains the derivative of the metric (gravitational potential) variable $\nu$ and a term that includes the local anisotropy distribution (a pressure difference divided by the radial coordinate) in Eq. (\ref{TOV}). On the other hand, $\mathcal{R}$ has dimensions of force per unit volume so, it is the total force per unit volume on each fluid element. Now, if the system is in equilibrium, these contributions cancel out so that $\mathcal{R}=0$ (a vanishing total force). Nevertheless, in the case of generating (via perturbations) a dynamic instability we will obtain a nonzero local contribution representing the hydrodynamic force on each fluid element. 
The scheme that is presented to perturb the system clearly produces a nonzero $\mathcal{R}$ immediately after taking the system out of equilibrium, which allows us to analyze the cracking of the stellar object.\\

For the physical variables appearing in (\ref{TOVb}) the following boundary conditions apply
\begin{eqnarray}
 m(0)=0,\qquad m(\Sigma)=M, \qquad P_{r}(r_{\Sigma})=0.
\end{eqnarray}
Polytropes  are static fluid configurations, which satisfy either  equation (\ref{p1})  or (\ref{p2}). The full set of equations describing the structure of these self-gravitating objects, in both cases, were derived and discussed in \cite{2p,3p,6p,7p}.

All the models have to satisfy physical requirements such as:
\begin{equation}
\rho >0, \qquad  \frac{P_r}{\rho}\leq 1, \qquad  \frac{P_\perp }{\rho} \leq 1.
\label{conditions}
\end{equation}

As already mentioned in the previous section, in order to integrate (\ref{TOV2}), we shall need an additional condition, besides (\ref{p2}). In this work such conditions, on the metric variables and their derivatives, are obtained by imposing that the only non-zero component of the Weyl tensor is zero (conformally flat condition) \cite{3p}, the class one (Karmarkar) condition \cite{7p} and the reasonable fact that we can choose for the tangential pressure also a polytrope equation of state \cite{6p}. Our last model considers that the polytrope has a zero complexity factor (vanishing complexity polytrope) \cite{VC}.

In what follows, we shall very briefly review the main equations corresponding to the general case.\\

\subsection{Relativistic polytrope for anisotropic matter}

This section is devoted to expose the basics of the theory of relativistic polytropes for anisotropic matter (for details see \cite{2p, 6p}). The starting assumption is to adopt the polytropic equation of state (\ref{p2}) for the radial pressure. As it is well known from the general theory of polytropes, there is a bifurcation at the value $\gamma=1$. Thus, the cases $\gamma=1$ and $\gamma\neq1$ have to be considered separately. In the context of our work we need bounded fluid configurations so the $\gamma=1$ will not be considered. 

Let us define the  variable $\Psi$ by
\begin{eqnarray}\label{rho2}
\rho=\rho_{c}\Psi^{n},
\end{eqnarray}
where $\rho_{c}$ denotes the energy density at the center (from now on the subscript $c$ indicates that the variable is evaluated at the center), and we express (\ref{p2}) as
\begin{eqnarray}\label{Pr}
P_{r}=K\rho_{c}^{\gamma}\Psi^{n \gamma}=P_{rc}\Psi^{1+n},
\end{eqnarray}
with $P_{rc}=K\rho_{c}^{\gamma}$. Note that from (\ref{Pr}), we can write
\begin{eqnarray}
P'_{r}=P_{rc}(1+n)\Psi^{n}\Psi ',
\end{eqnarray}
so that the TOV equation (\ref{TOV}) can be written as
\begin{eqnarray}
2\alpha(1+n)\Psi '+(\alpha\Psi +1)\nu'-4\frac{\Delta}{r\rho_{c}\Psi^{n}}=0,
\end{eqnarray}
where we have defined $\alpha=\frac{P_{rc}}{\rho_{c}}$. From this we get
\begin{eqnarray}\label{nup}
\nu'=\frac{4\Delta}{r\rho_{c}\Psi^{n}(
\alpha \Psi +1)}-\frac{2\alpha (1+n)}{\alpha \Psi +1}\Psi '\label{nup}.
\end{eqnarray}
The integration of (\ref{nup}), and the subsequent replacement of (\ref{mf}) and (\ref{nup}) in (\ref{ee2}), produces
\begin{eqnarray}
&&\alpha w\frac{dm}{dr}+\frac{m}{r}+\alpha (1+n)\frac{r}{1+\alpha w}\frac{dw}{dr}\left(1-\frac{2m}{r}\right)\nonumber\\
&&-\frac{2\Delta}{\rho_{c}}\frac{(1-\frac{2m}{r})}{(1+\alpha w)w^{n}}=0.\label{le0}
\end{eqnarray}

Let us now introduce the following dimensionless variables
\begin{eqnarray}
\eta&=&\frac{m(r)A^3}{4\pi \rho_{c}},\label{d1}\\
r&=&\xi/A,\label{d2}\\
A^{2}&=&
\frac{4\pi \rho_{c}}{(n+1)\alpha} \label{d3},
\end{eqnarray}
in terms of which (\ref{le0}) can be written as 
\begin{eqnarray}\label{lemd}
&&\left[\frac{\xi -2\alpha (1+n)\eta}{1+\alpha \Psi }\right]
\left[\xi \Psi '-\frac{2\Delta}{\rho_{c}\alpha (1+n)\Psi ^{n}}\right]\nonumber\\
&&+\eta+\alpha \xi \Psi \eta'=0,
\end{eqnarray}
where (see reference \cite{6p} for details)
\begin{eqnarray}\label{lemd2}
\eta'=\xi^{2}\Psi^{n}.
\end{eqnarray}
Notice that from now on the prime denotes derivative with respect to the variable $\xi$. At the boundary surface $r=r_\Sigma$ ($\xi=\xi_\Sigma$) we have $\Psi (\xi_\Sigma)=0$. In this case, conditions (\ref{conditions}) read:
\begin{equation}
\rho>0, \qquad \alpha \Psi \leq 1, \qquad \alpha \Psi + \frac{\Delta}{\rho_c \Psi^n} \leq1.
\label{conditionsIII}
\end{equation}

Equations (\ref{lemd}) and (\ref{lemd2}), form a system of two first order ordinary differential equations for the three unknown functions: $\Psi, \eta, \Delta$, depending on a duplet of parameters $n, \alpha$. Thus, as we have mentioned, it is obvious that in order to proceed further with the modeling of a compact object, we need to provide additional information. Such  information, of course, depends on the specific physical problem under consideration. As we have already mentioned here, we shall further assume the metric conditions or equations of state used in the models built up in \cite{3p, 6p, 7p, VC}.

\section{Perturbation scheme}
 In this work, we propose a schematic perturbation scheme which corresponds to an extension of \cite{cn4}. Let us start with a spherical anisotropic relativistic fluid distribution satisfying the generalized hydrostatic equilibrium equation (\ref{TOV}). Besides, the pressures are considered as functions of the energy density and the anisotropic function, i.e.  
\begin{equation}
P_r(\rho,\Delta), \quad P_{\perp}(\rho,\Delta).
\end{equation}
Now, to study the appearance of cracking, we shall perform perturbations of the energy density and the local pressure anisotropy 
\begin{eqnarray}
    \tilde{\rho} &=& \rho + \delta \rho, \\
    \tilde{\Delta} &= & \Delta + \delta \Delta,
\end{eqnarray}
where $\delta \rho$ and $\delta \Delta$ indicate small perturbations that may depend on $r$. Thus, we can write the perturbed quantities (up to first order) like,
\begin{eqnarray}
    P_r \; &\longrightarrow& \; \tilde{P}_r = P_r + \left(\frac{\partial\tilde{P}_r}{\partial \tilde{\rho}}\right)_{\begin{array}{c}
         \tilde{\rho}=\rho  \\
         \tilde{\Delta}= \Delta 
    \end{array}} \delta\rho \nonumber \\ && \hspace{0.5cm}+\left(\frac{\partial\tilde{P}_r}{\partial \tilde{\Delta}}\right)_{\begin{array}{c}
         \tilde{\rho}=\rho  \\
         \tilde{\Delta}= \Delta 
    \end{array}} \delta\Delta,  \\ && \nonumber \\
    m\; &\longrightarrow& \; \tilde{m} = m + \left(\frac{\partial\tilde{m}}{\partial \tilde{\rho}}\right)_{\begin{array}{c}
         \tilde{\rho}=\rho   \\
         \tilde{\Delta}= \Delta 
    \end{array}} \delta\rho, \\
    \Delta \; &\longrightarrow&\; \tilde{\Delta} = \Delta + \delta \Delta.
\end{eqnarray}
Now, let us assume
\begin{eqnarray}\label{funcionP}
    \tilde{P}_r = (1+\delta \phi) P_r \; , \quad |\delta \phi|<<1,
\end{eqnarray}
where $\delta \phi$ is a constant which ensures that the radial pressure maintains the same functional behaviour. As a consequence, we have
\begin{eqnarray}
   \frac{d\tilde{P}_r}{dr} = (1+\delta \phi) \ \frac{dP_r}{dr} \quad \Longrightarrow \quad \delta P'_r = P'_r \ \delta \phi. 
\end{eqnarray}
Moreover, this implies a restriction over the perturbation functions, which is 
\begin{eqnarray*}
   && \left(\frac{\partial\tilde{P}_r}{\partial \tilde{\rho}}\right)_{\begin{array}{c}
         \tilde{\rho}=\rho  \\
         \tilde{\Delta}= \Delta 
    \end{array}} \delta \rho + \left(\frac{\partial\tilde{P}_r}{\partial \tilde{\Delta}}\right)_{\begin{array}{c}
         \tilde{\rho}=\rho  \\
         \tilde{\Delta}= \Delta 
    \end{array}} \delta \Delta = P_r \ \delta \phi ,
\end{eqnarray*}
which leads to 
\begin{eqnarray}
   \delta \rho &=& \left\lbrace \left(\frac{\partial\tilde{P}_r}{\partial \tilde{\rho}}\right)^{-1} \left(P_r \delta \phi - \left(\frac{\partial \tilde{P}_r}{\partial \tilde{\Delta}}\right)\delta \Delta\right) \right\rbrace_{\begin{array}{c}
         \tilde{\rho}=\rho \\
         \tilde{\Delta}= \Delta 
    \end{array}}.
\end{eqnarray}
and in this way (\ref{funcionP}) is satisfied.

The total radial force of the gravitational system before the perturbation is defined, by means of (\ref{TOV2}), as
\begin{eqnarray}
\mathcal{R}\equiv P_r' + \frac{m+4\pi r^3 P_r}{r(r-2m)}(\rho + P_r) -2\frac{\Delta}{r}=0 .  \label{eq}
\end{eqnarray}
Then, after perturbation  we can write 
\begin{eqnarray}
   \tilde{\mathcal{R}}(\tilde{\rho},\tilde{\Delta}) = \mathcal{R}(\rho,\Delta)+\delta \mathcal{R} (\rho,\Delta),
\end{eqnarray}
where
\begin{eqnarray}
   \delta \mathcal{R} &=& \left(\frac{\partial \mathcal{R}}{\partial P_r}\right)_{\begin{array}{c}
         \tilde{\rho}=\rho  \\
         \tilde{\Delta}= \Delta 
    \end{array}}\delta P_r + \left(\frac{\partial \mathcal{R}}{\partial \rho}\right)_{\begin{array}{c}
         \tilde{\rho}=\rho  \\
         \tilde{\Delta}= \Delta 
    \end{array}}\delta \rho \nonumber \\ &+& \left(\frac{\partial \mathcal{R}}{\partial m}\right)_{\begin{array}{c}
         \tilde{\rho}=\rho  \\
         \tilde{\Delta}= \Delta 
    \end{array}} \delta m \nonumber + \left(\frac{\partial \mathcal{R}}{\partial \Delta}\right)_{\begin{array}{c}
         \tilde{\rho}=\rho  \\
         \tilde{\Delta}= \Delta 
    \end{array}}\delta \Delta \nonumber \\ &+& \left(\frac{\partial \mathcal{R}}{\partial P'_r}\right)_{\begin{array}{c}
         \tilde{\rho}=\rho  \\
         \tilde{\Delta}= \Delta 
    \end{array}} \delta P'_r.
\end{eqnarray}
From Eq. (\ref{eq}), it is straightforward to show that
\begin{eqnarray}
\frac{\partial \mathcal{R}}{\partial P_r} &=& \frac{4\pi r}{1-2m/r}(\rho + P_r) + \frac{m+4\pi r^3 P_r}{r^2(1-2m/r)}, \\
\frac{\partial \mathcal{R}}{\partial \rho} &=& \frac{m+4\pi r^3 P_r}{r^2(1-2m/r)}, \\
\frac{\partial \mathcal{R}}{\partial \Delta} &=& -\frac{2}{r}, \\
\frac{\partial \mathcal{R}}{\partial P'_r} &=& 1, \\
\frac{\partial \mathcal{R}}{\partial m} &=& \frac{\rho+ P_r}{(r-2m)^2}(1+8\pi r^2P_r).
\end{eqnarray}
Now, for simplicity, let us write 
\begin{eqnarray}
\delta \Delta = f(r)\delta\beta\;, \quad |\delta \beta|<<1,
\end{eqnarray}
where $\delta \beta$ is constant and $f(r)$, is arbitrary. However, to ensure that the radial force in the center of the distribution remains finite we will assume that
\begin{eqnarray}\label{fc}
\lim_{r\rightarrow 0} \frac{f(r)}{r} = 0.
\end{eqnarray}
Thus, using (\ref{mf2}) we obtain
\begin{eqnarray}
\delta m = 4\pi (F_1(r) \delta \phi - F_2(r) \delta \beta),
\end{eqnarray}
with
\begin{eqnarray}
F_1(r) &\equiv& \int^r_0 \Bar{r}^2  G(r) P_r d\Bar{r}, \\
F_2(r) &\equiv& \int^r_0 \Bar{r}^2 G(r) \left(\frac{\partial \tilde{P}_r}{\partial \tilde{\Delta}}\right) f(r) d\Bar{r}, \\
G(r) &\equiv& \left(\frac{\partial\tilde{P}_r}{\partial \tilde{\rho}}\right)^{-1}\; .
\end{eqnarray}
Finally, the total radial force after the perturbation 
reads
\begin{eqnarray}\label{Rfinal}
\mathcal{\tilde{R}} &=& \Bigg\{P_r \Bigg[\frac{4\pi r(\rho+ P_r)}{1-2m/r}+(1+G(r))\Bigg(\frac{m+4\pi r^3P_r}{r^2(1-2m/r)}\Bigg)\Bigg] \nonumber \\ &+& \frac{4\pi(\rho+P_r)(1+8\pi r^2 P_r)F_1(r)}{r^2(1-2m/r)^2} +  P'_r\Bigg\}\;\delta \phi \nonumber \\ &-& \Bigg\{ G(r)f(r)\Bigg[\Bigg(\frac{\partial P_r}{\partial \Delta}\Bigg)\Bigg(\frac{m+4\pi r^3P_r}{r^2(1-2m/r)}\Bigg) \Bigg] \nonumber \\ &+& \frac{4\pi(\rho+P_r)(1+8\pi r^2 P_r)F_2}{r^2(1-2m/r)^2} + \frac{2}{r}f(r)\Bigg\} \;\delta \beta .
\end{eqnarray}
Now, it is clear that the change of sign  which has to be present in the total radial force, required for the existence of cracking (or overturning), implies $\mathcal{R} =0$ for some $ r \in (0,r_{\Sigma})$. This leads to
\begin{eqnarray}\label{Gamma}
\delta \phi =  \Gamma \delta \beta, 
\end{eqnarray}
where
\begin{eqnarray}
\Gamma^{-1}  &=& \Bigg\{P_r \Bigg[\frac{4\pi r(\rho+ P_r)}{1-2m/r}+(1+G(r))\Bigg(\frac{m+4\pi r^3P_r}{r^2(1-2m/r)}\Bigg)\Bigg] \nonumber \\ &+& \frac{4\pi(\rho+P_r)(1+8\pi r^2 P_r)F_1(r)}{r^2(1-2m/r)^2}  + P'_r\Bigg\} \Bigg/ \Bigg\{ G(r)f(r) \nonumber \\ &\times&\Bigg[\Bigg(\frac{\partial P_r}{\partial \Delta}\Bigg)\Bigg(\frac{m+4\pi r^3P_r}{r^2(1-2m/r)}\Bigg) \Bigg] \nonumber \\& + &\frac{4\pi(\rho+P_r)(1+8\pi r^2 P_r)F_2}{r^2(1-2m/r)^2} +  \frac{2}{r}f(r)\Bigg\}. \label{crac}
\end{eqnarray}

At this point, a couple of comments are in order. First, note that with equations (\ref{Rfinal})-(\ref{crac}) is possible to evaluate the occurrence of cracking (overturning) in any spherically symmetric system satisfying a barotropic/polytropic equations of state. Furthermore, if the system satisfies the physical acceptability conditions, it is easy to show that the total radial force will be free of singularities and equal to zero in the center of the distribution. Second, when a perturbation is introduced using our scheme, the modified (anisotropic) TOV equation does not vanish anymore. The rest of the work is devoted to the implementation of the scheme developed here in some particular models.


\section{Models}
\label{conc10ch10}

In this section we shall study the appearance of cracking (overturning) for systems described by a polytropic equation of state given by Eq. (\ref{p2}). In this case, the radial pressure is function of the energy density only which implies a huge simplification in the perturbation method. Also, the parameter $\delta \phi$ measures the perturbation in the energy density while $\delta \beta$ measures the perturbation over the local anisotropy. Now, for simplicity we will consider that $f(r)=\Delta$. Evidently, for a well behaved internal solution, this choice satisfies (\ref{fc}). Then, using Eq. (\ref{rho2}) and the dimensionless variables presented in Eqs. (\ref{d1})-(\ref{d3}), we obtain directly from (\ref{Rfinal}) the total radial force as 
\begin{eqnarray}
\mathcal{\hat{R}} &=& \Bigg\{ \Psi^{n} \frac{d\Psi}{d\xi}+\frac{\alpha \Psi^{n+1}}{a(\xi)}\Bigg[\xi \Psi^{n}b(\xi) \nonumber \\ &+& \frac{c(\xi)}{\xi^2}\Bigg(1+\frac{n}{\alpha(n+1)\Psi}\Bigg)\Bigg] + \frac{\alpha n\Psi^{n}\eta}{\xi^2(n+1)}\frac{b(\xi)d(\xi)}{a(\xi)^2}\Bigg\} \;\delta\phi \nonumber \\ &-& \frac{2\Delta}{\alpha (n+1) \rho_c \xi}\; \delta \beta , \label{trf}
\end{eqnarray}
where
\begin{eqnarray}
\mathcal{\hat{R}} &\equiv& \left(\frac{A}{4\pi \rho_c^2}\right)\tilde{R}, \\
a(\xi) &\equiv& 1-2(n+1)\eta/\xi, \\
b(\xi) &\equiv& 1+\alpha \Psi, \\
c(\xi) &\equiv& \eta + \alpha \xi^3 \Psi^{n+1}, \\
d(\xi) &\equiv& 1+2\xi^2(n+1)\alpha^2 \Psi^{n+1}.
\end{eqnarray}

At this point it should be emphasized that evaluating $\hat{\mathcal{R}}$ in Eq. (\ref{trf}) requires the knowledge of the pair $\{\psi,\eta\}$ for certain model specified by the anisitropy $\Delta$. Nevertheless, the generalized Lane--Emden system, Eqs. (\ref{lemd}) and (\ref{lemd2}), does not admit analytical solution so all our computations here will be numerically performed. To be more precise, we should integrate numerically for the pair $\{\psi,\eta\}$ and then use the values to evaluate $\hat{\mathcal{R}}$. In what follows we shall consider some particular cases.

\subsection{The conformally flat polytrope}

The conformally flat polytrope is obtained by imposing the condition for the vanishing of the Weyl tensor $C_{\mu \nu \rho \lambda}$. In the spherically symmetric case all components of Weyl tensor become proportional to a single scalar function, expressed through the component $C^3_{232}$,  
\begin{eqnarray}
W & \equiv & \frac{r}{2}C^3_{232}\nonumber \\ 
&=&\frac{r^3e^{-\lambda}}{6}\Bigg( \frac{e^{\lambda}}{r^2}+\frac{\nu'\lambda'}{4}-\frac{1}{r^2}- \frac{\nu'^2}{4}-\frac{\nu''}{2} - \frac{\lambda'-\nu'}{2r} \Bigg). \nonumber\\
\end{eqnarray}
Then, it can be established a relation that expresses the Weyl tensor in terms of the energy density contrast and the local anisotropy of pressure (see \cite{Herrera2001}),  
\begin{eqnarray}
W = - \frac{4\pi}{3}\int_{0}^{r} r^3 \rho ' dr + \frac{4\pi}{3} r^3 (P_r - P_{\perp}).
\end{eqnarray}
It has been shown in Ref. \cite{Herrera2001} that the conformally flat condition ($W = 0$) can be integrated, producing 
\begin{eqnarray}\label{W=0}
e^{\nu} = B^2 r^2 cosh^{2}\left[\int \frac{e^{\lambda/2}}{r} dr + A \right].
\end{eqnarray}
where $B$ and $A$ are integration constants. Thus, the conformally flat condition reduces the number of unknown functions of the system. Using the field equations (\ref{ee2}), (\ref{ee3}) and the condition $W=0$ we get an anisotropic function that can be written as
\begin{eqnarray}
\Delta = \frac{r}{8\pi}\left(\frac{e^{-\lambda}-1}{r^2}\right)'.
\end{eqnarray}
Then, in the case represented by (\ref{p2})  we obtain 
\begin{eqnarray}
\Delta = \rho_c \left(\frac{3\eta}{\xi^3}-\Psi^n\right).
\end{eqnarray}

\begin{figure}[ht!]
\centering
\includegraphics[scale=0.5]{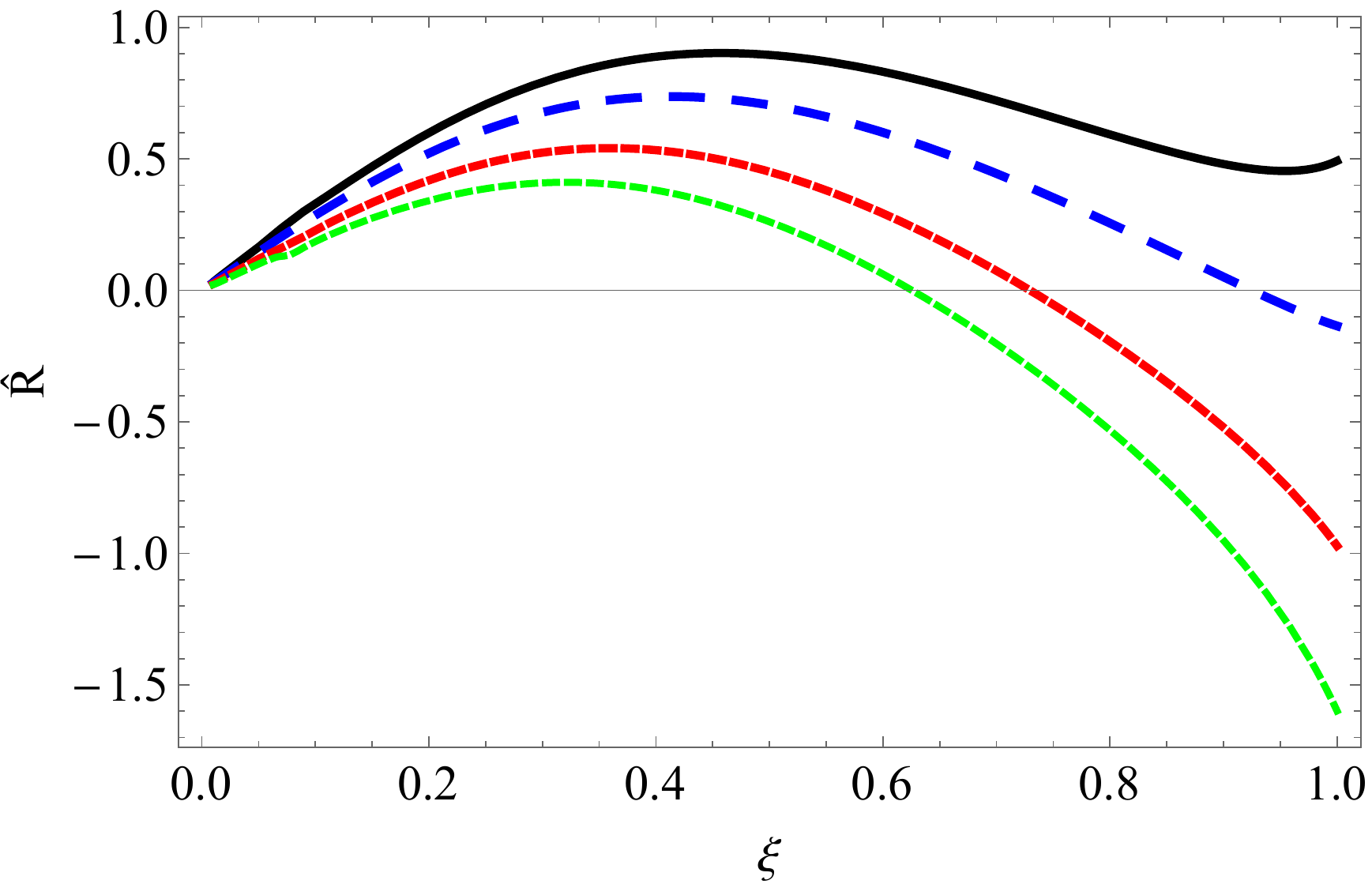}
\caption{\label{c-cf-gamma-var}
$\mathcal{\hat{R}}$ as function of $\xi/\xi_{\Sigma}$, for 
$n=0.1$, $\alpha=1$,  and $\Gamma=-5$ (black line), $\Gamma=-2$ (blue line), $\Gamma=2$ (red line) and $\Gamma=5$ (green line)
}
\end{figure}

\begin{figure*}[ht!]
\centering
\includegraphics[width=0.4\textwidth]{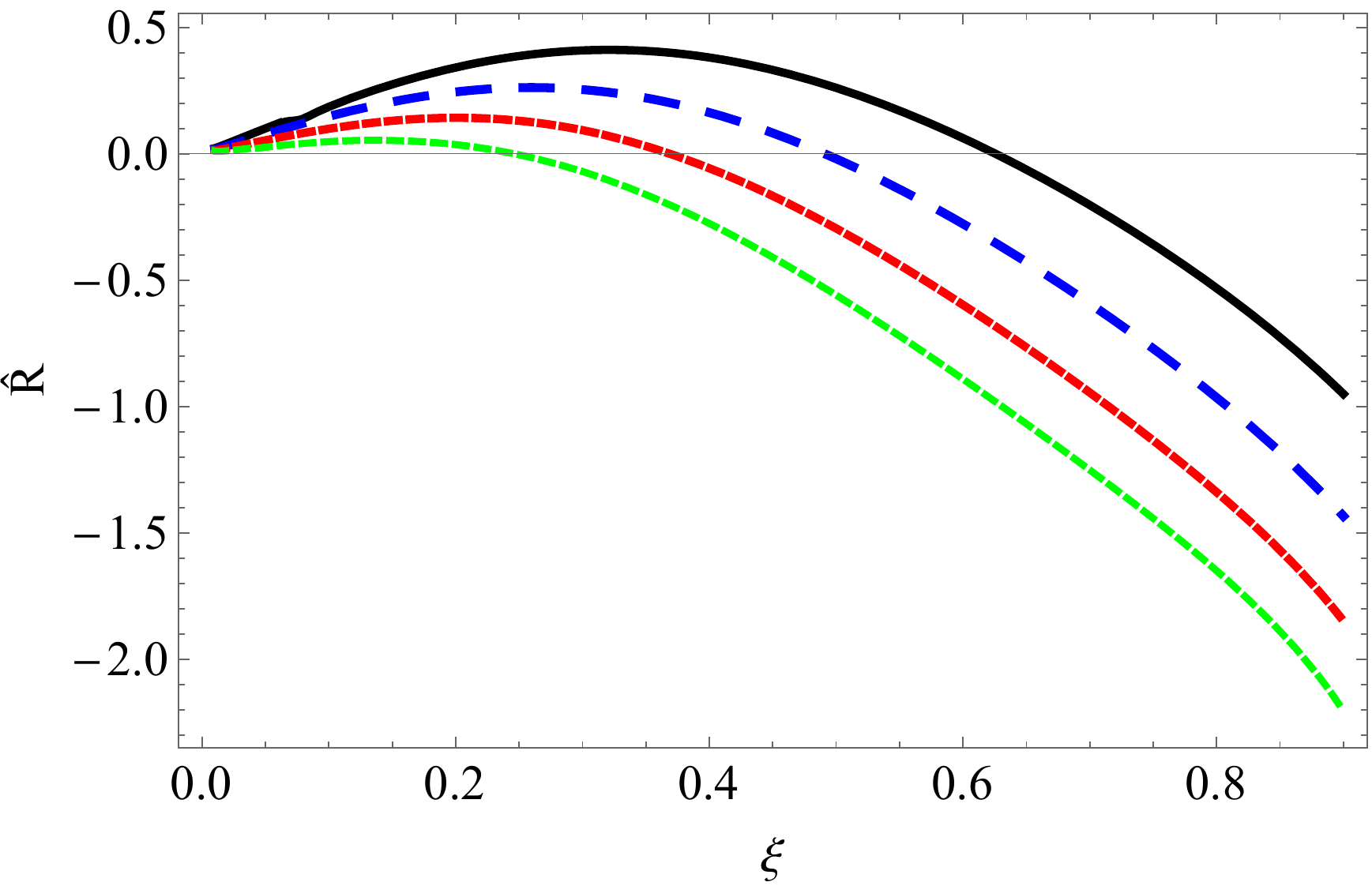}  \
\includegraphics[width=0.4\textwidth]{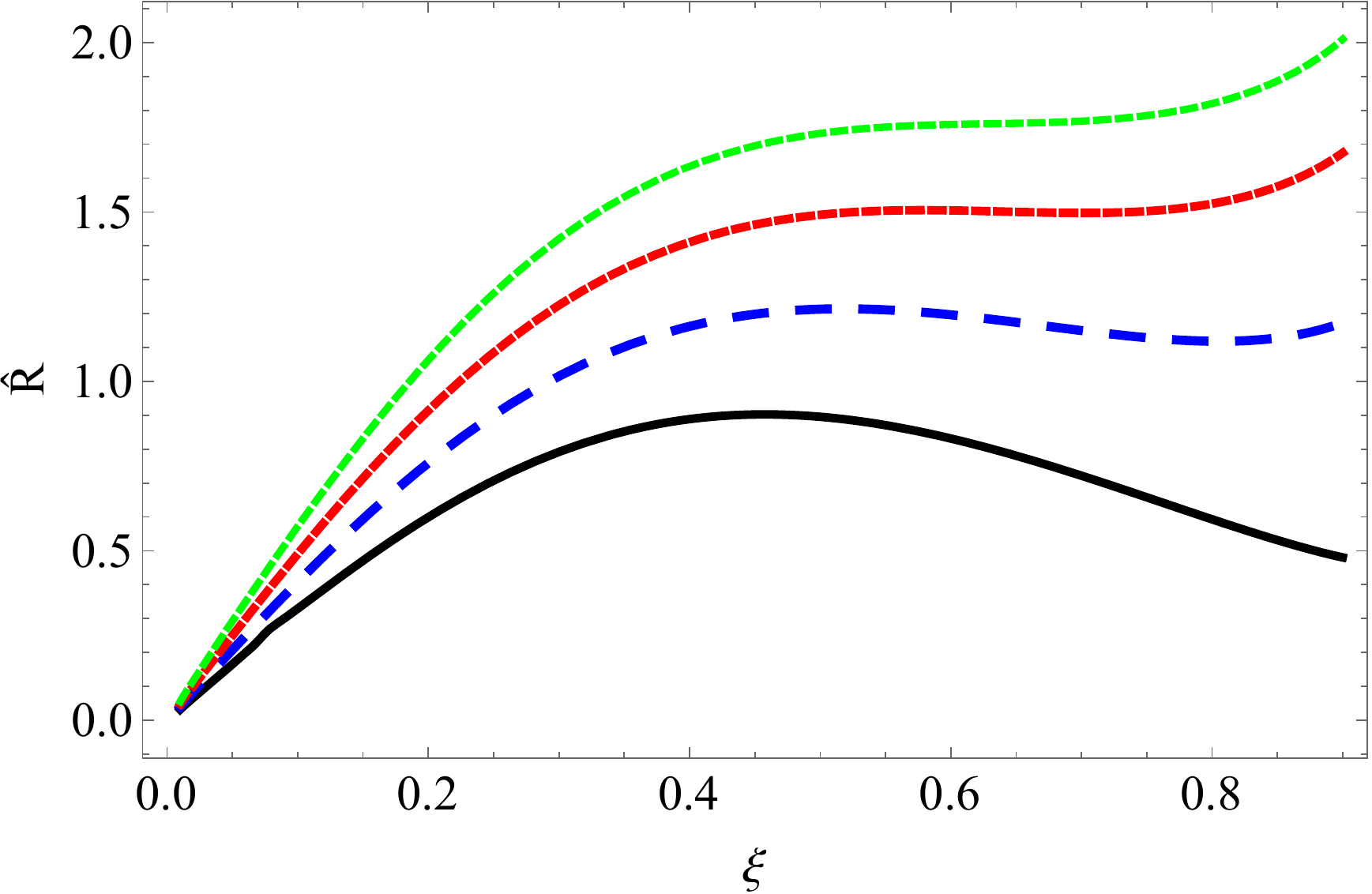}  \
\caption{\label{c-cf-n-var}
$\mathcal{\hat{R}}$ as function of $\xi/\xi_{\Sigma}$, for 
$\alpha=1$, $n=0.1$ (black line), $n=0.2$ (blue line), $n=0.3$ (red line) and $n=0.4$ (green line)} 
with $\Gamma=5$ (left panel) and $\Gamma=-5$ (right panel)
\end{figure*}

\begin{figure*}[ht!]
\centering
\includegraphics[width=0.4\textwidth]{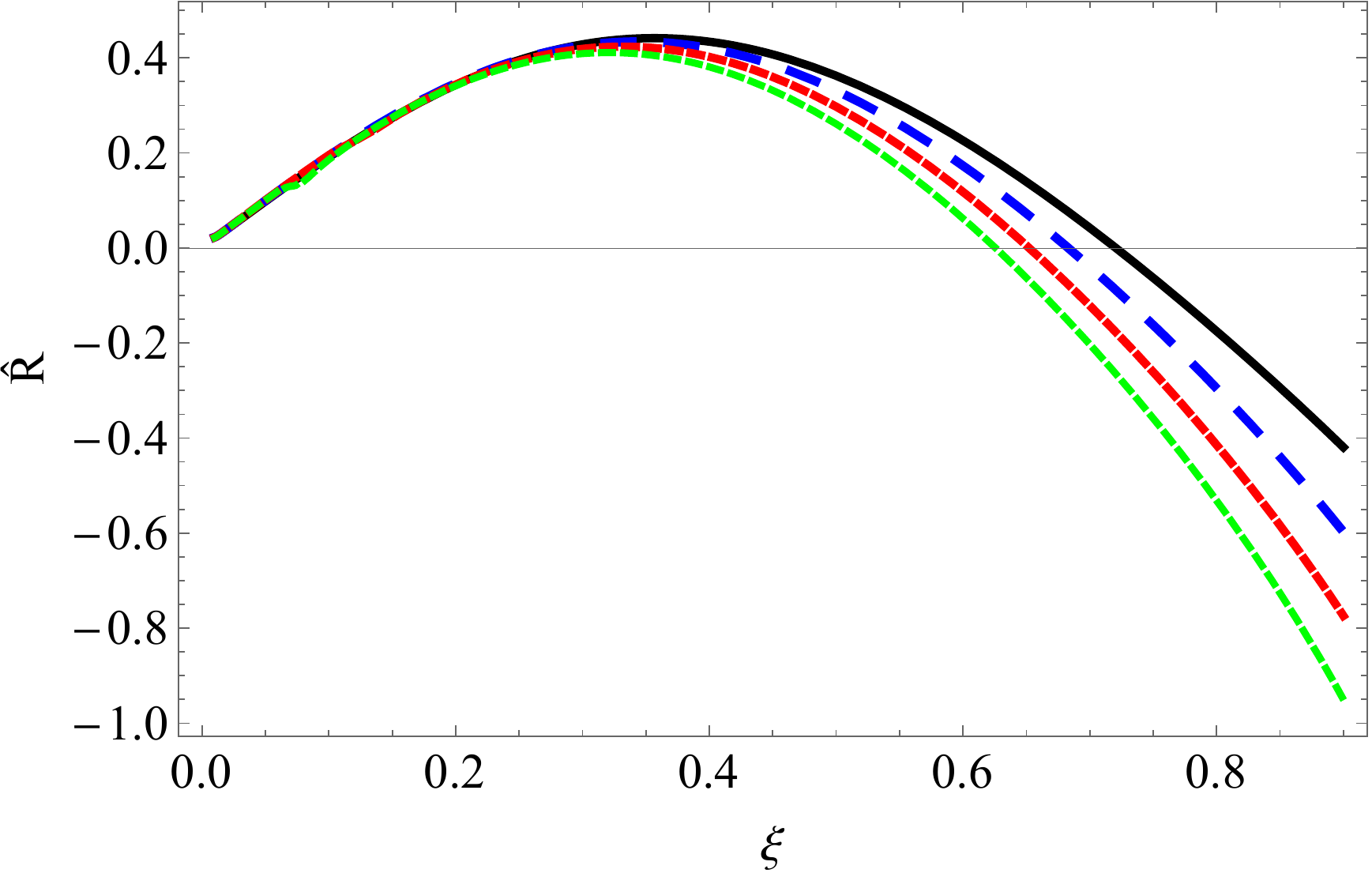}  \
\includegraphics[width=0.4\textwidth]{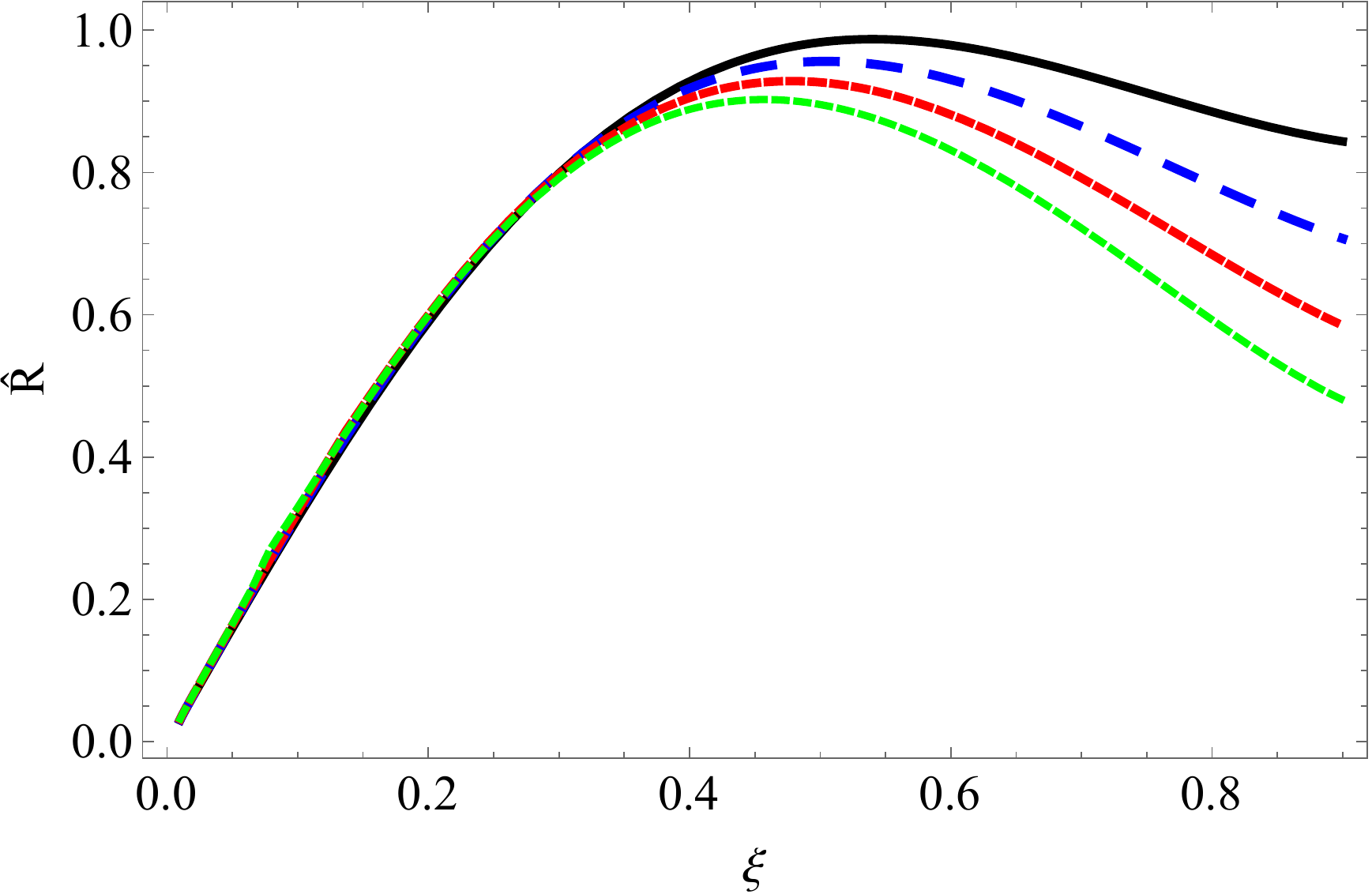}  \
\caption{\label{c-cf-alfa-var}$\mathcal{\hat{R}}$ as function of $\xi/\xi_{\Sigma}$, for 
$n=0.1$, $\alpha=0.7$ (black line), $\alpha=0.8$ (blue line), $\alpha=0.9$ (red line) and $\alpha=1$ (green line) with $\Gamma=5$ (left panel) and $\Gamma=-5$ (right panel)}
\end{figure*}

\begin{figure}[ht!]
\centering
\includegraphics[scale=0.5]{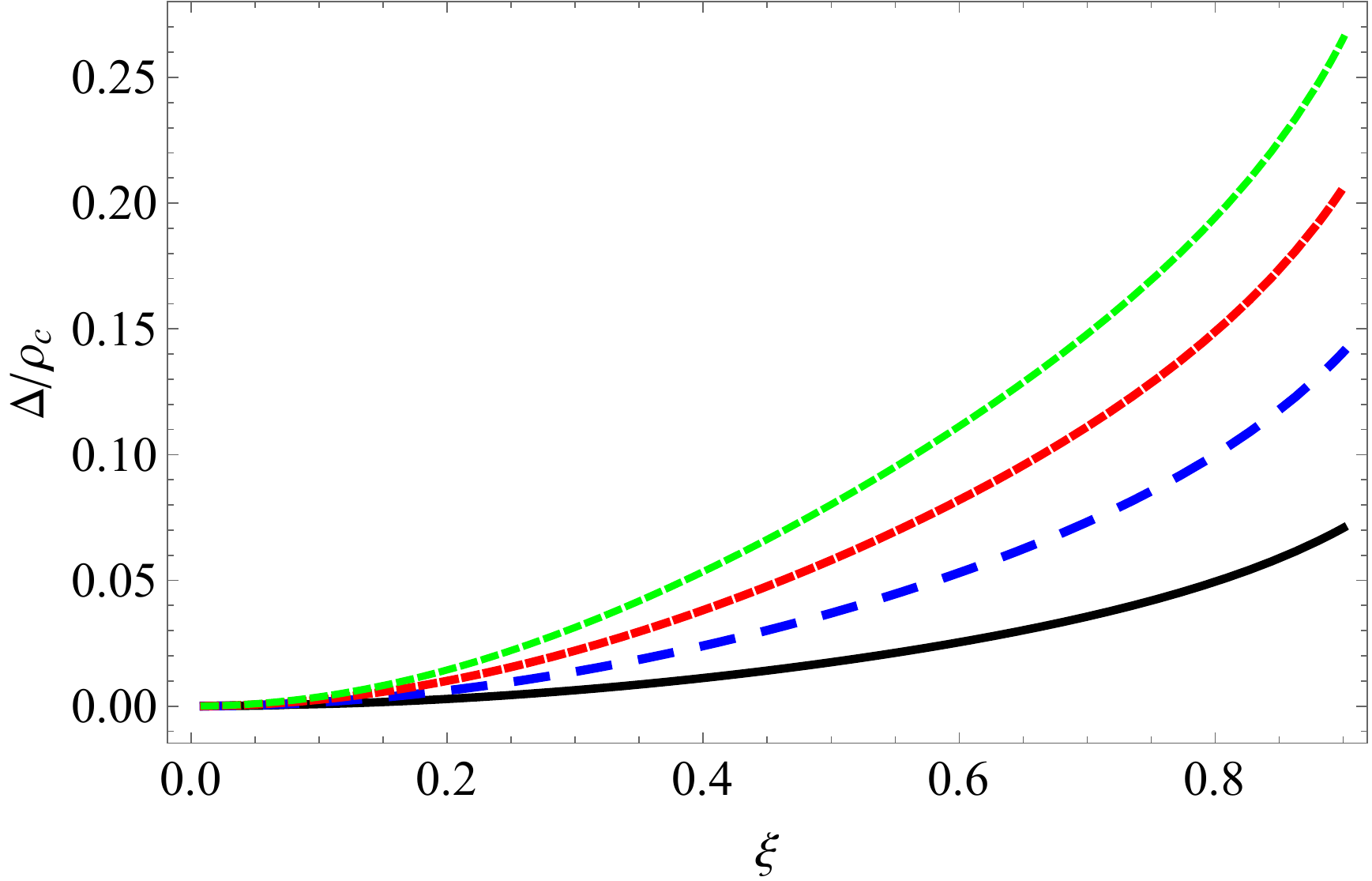}
\caption{\label{a-cf-n-var}
Anisotropy $\Delta/\rho_{c}$ as function of $\xi/\xi_{\Sigma}$, for $\alpha=1$ and $n=0.1$ (black line), $n=0.2$ (blue line), $n=0.3$ (red line) and $n=0.4$ (green line)
}
\end{figure}

Now, we are ready to write the total radial force (\ref{trf}) like
\begin{eqnarray}
\mathcal{\hat{R}} &=& \Bigg\{ \Psi^{n} \frac{d\Psi}{d\xi}+\frac{\alpha \Psi^{n+1}}{a(\xi)}\Bigg[\frac{c(\xi)}{\xi^2}\Bigg(1+\frac{n}{\alpha(n+1)\Psi}\Bigg) \nonumber \\ &+& \xi \Psi^{n}b(\xi)\Bigg] + \frac{\alpha n\Psi^{n}\eta}{\xi^2(n+1)}\frac{b(\xi)d(\xi)}{a(\xi)^2}\Bigg\} \delta\phi \nonumber \\ &-& \frac{2}{\alpha (n+1) \xi}  \left(\frac{3\eta}{\xi^3}-\Psi^n\right) \delta \beta . \label{rfcf}
\end{eqnarray}
In Figs. \ref{c-cf-gamma-var}, \ref{c-cf-n-var} and \ref{c-cf-alfa-var} we show the behaviour of $\mathcal{\hat{R}}$ for different values of the parameters involved. To complement the discussion, in Fig. \ref{a-cf-n-var}, we show the behaviour of the anisotropy as a function of the polytropic index $n$.

\subsection{The Karmarkar class I polytrope}

Embedding of four-dimensional spacetimes into higher dimensions is an invaluable tool in generating both cosmological and astrophysical models \cite{maurya}. Recently, in reference \cite{7p} was exposed in detail a model for a compact object composed by an anisotropic fluid, that meets the equation of state of the polytrope together with the Karmarkar condition. We will summarize the basic aspects that lead us to obtain a particular anisotropy for this model. As it is well known, the Karmarkar condition is necessary and sufficient to ensure class one solutions \cite{karmarkar} which  for spherically symmetric space--times reads ($R_{\theta\phi\theta\phi}\neq0$)
\begin{equation}\label{eq15}
R_{r t r t}R_{\theta\phi \theta\phi}=R_{r\theta r \theta}R_{t \phi t \phi}+R_{r\theta t \theta}R_{r\phi t\phi},    
\end{equation}
leading to
\begin{equation}\label{eq16}
2\frac{\nu^{\prime\prime}}{\nu^{\prime}}+\nu^{\prime}=\frac{\lambda^{\prime}e^{\lambda}}{e^{\lambda}-1},
\end{equation}
with $e^{\lambda}\neq 1$.
Now,  using (\ref{rho2}), (\ref{Pr}) and (\ref{eq16}) in (\ref{delta}) we obtain
\begin{eqnarray}\label{anis:dimen}
 \Delta = \frac{ \left( 4 \pi P_{rc} \Psi^{n+1} r^3 - m \right)\left(r m' - 3 m \right)  }{16 \pi m\ r^3 },
\end{eqnarray}
from where, after using the set of dimensionless variables (\ref{d1}), (\ref{d2}) and (\ref{d3}), we arrive at
\begin{eqnarray}\label{anisC1}
\Delta = \rho_c \frac{\left( \alpha \Psi^{n+1} \xi^3 -  \eta \right)\left(\xi^3 \Psi^{n} - 3 \eta   \right)}{ 4 \xi^3 \eta},
\end{eqnarray}
which represents the particular form of the anisotropy obtained for the class one polytrope. Then, after we perturb the system, the total radial force is given by,
\begin{eqnarray}
\mathcal{\hat{R}} &=& \Bigg\{ \Psi^{n} \frac{d\Psi}{d\xi}+\frac{\alpha \Psi^{n+1}}{a(\xi)}\Bigg[\frac{c(\xi)}{\xi^2}\Bigg(1+\frac{n}{\alpha(n+1)\Psi}\Bigg) \nonumber \\ &+& \xi \Psi^{n}b(\xi)\Bigg] + \frac{\alpha n\Psi^{n_r}\eta}{\xi^2(n+1)}\frac{b(\xi)d(\xi)}{a(\xi)^2}\Bigg\} \;\delta\phi \nonumber \\ &-& \frac{\left( \alpha \Psi^{n+1} \xi^3 -  \eta \right)\left(\xi^3 \Psi^{n} - 3 \eta   \right)}{ 2 \alpha (n+1) \xi^4  \eta}\; \delta \beta .
\end{eqnarray}

\begin{figure}[ht!]
\centering
\includegraphics[scale=0.5]{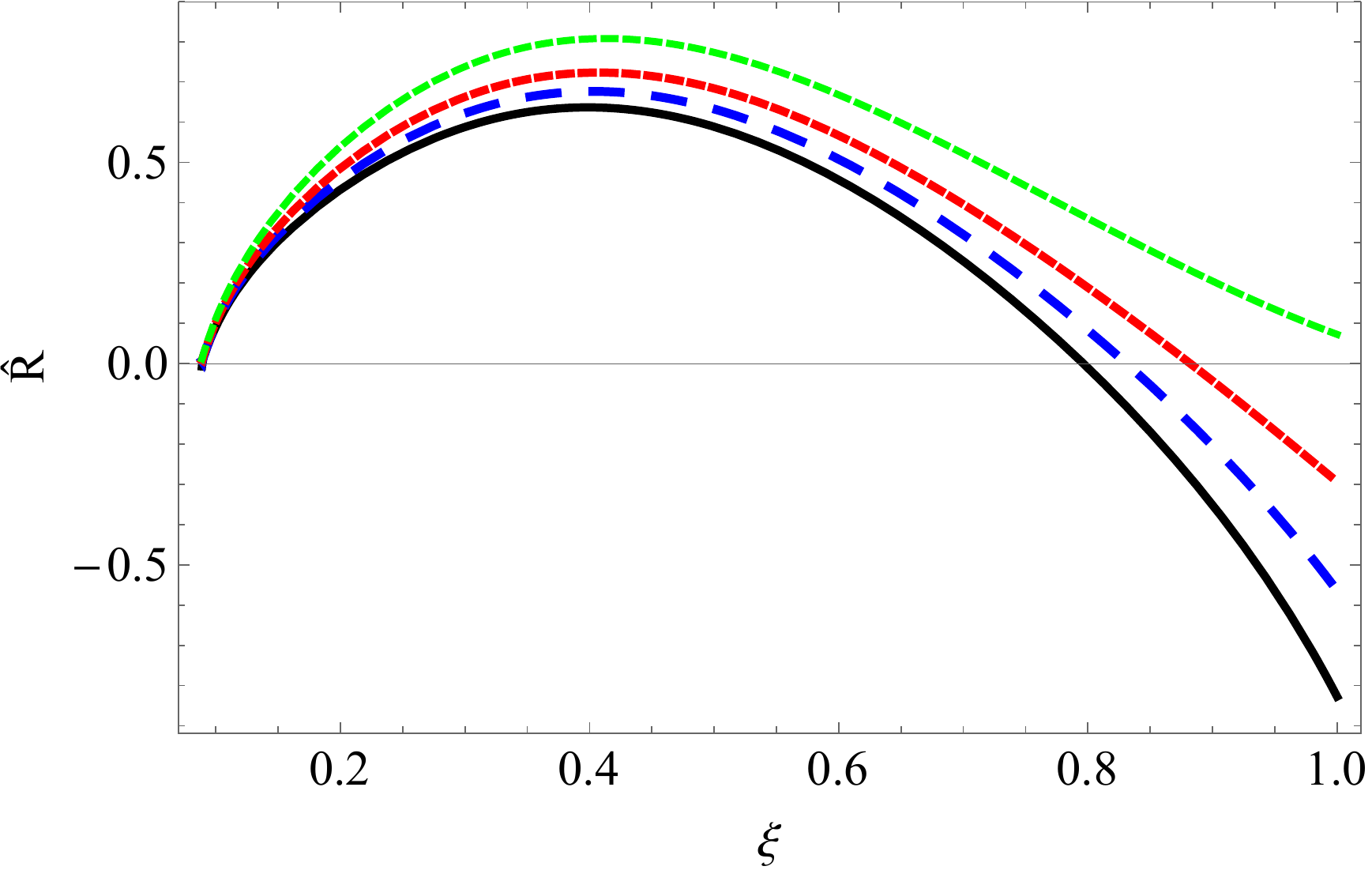}
\caption{\label{c-k-n-var}
$\mathcal{\hat{R}}$ as function of $\xi/\xi_{\Sigma}$, for 
$\alpha=1$, $\Gamma=1$ and $n=0.01$ (black line), $n=0.05$ (blue line), $n=0.1$ (red line) and $n=0.2$ (green line)
}
\end{figure}

\begin{figure}[ht!]
\centering
\includegraphics[scale=0.5]{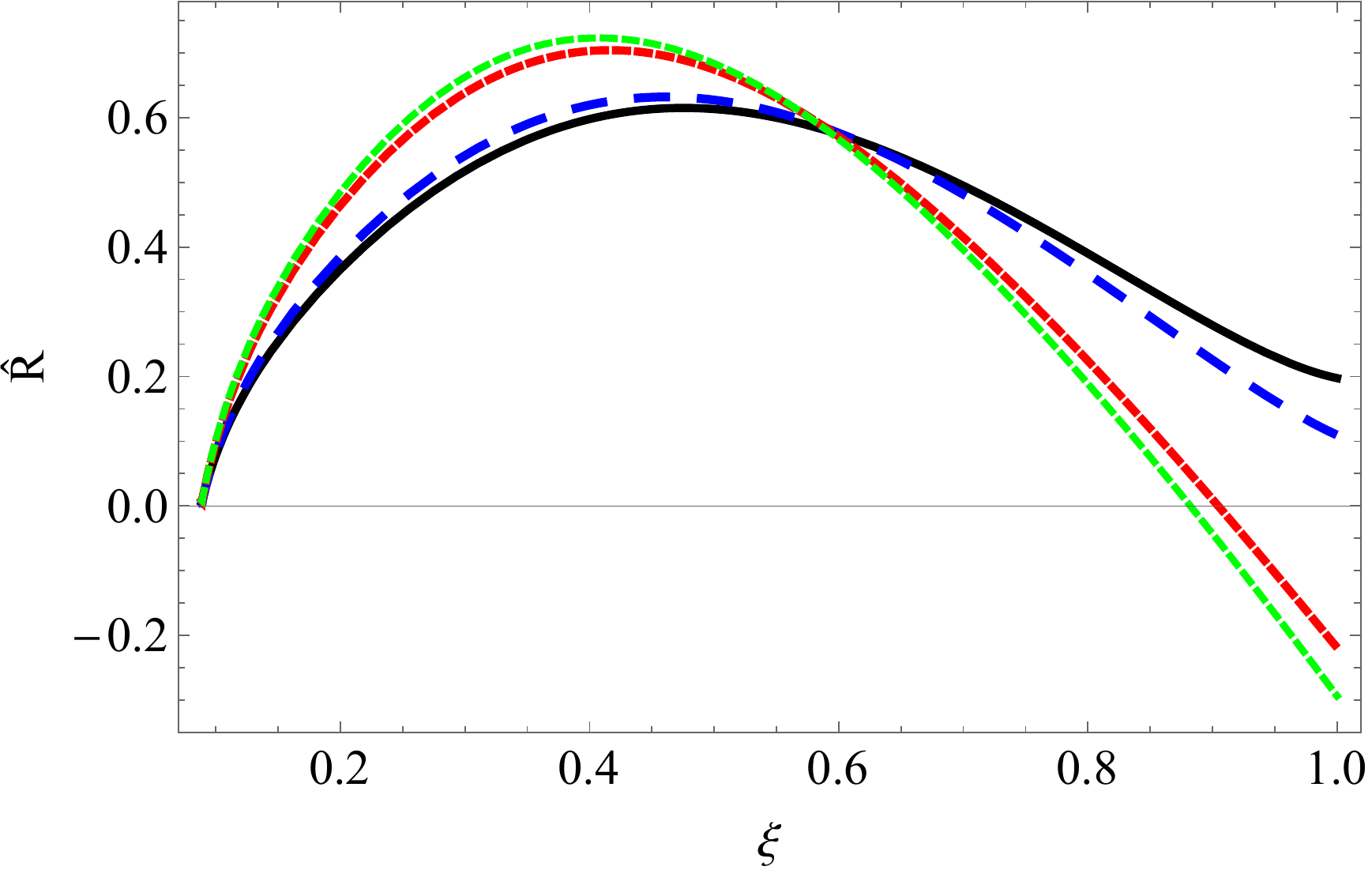}
\caption{\label{c-k-alfa-var}
$\mathcal{\hat{R}}$ as function of $\xi/\xi_{\Sigma}$, for 
$n=0.1$, $\Gamma=1$ and $\alpha=0.88$ (black line), $\alpha=0.9$ (blue line), $\alpha=0.98$ (red line) and $\alpha=1$ (green line)
}
\end{figure}

In Figs. \ref{c-k-n-var} and \ref{c-k-alfa-var} it is shown the behaviour of $\mathcal{\hat{R}}$ for different values of the parameters involved.

\subsection{The double polytrope}
In a recent paper \cite{6p} the extra information needed to close the system is supplied by the assumption that the tangential pressure also satisfies a polytropic equation of state. Since we have now two polytropic equations of state we shall clearly differentiate two polytropic exponents (indexes) $\gamma_r$, $\gamma_\bot$ ($n_r$, $n_\bot$), one  for each polytrope. Thus, three possible cases may be considered, but only two of these represent bounded configurations for the fluid distribution of the compact object. We will briefly expose these two cases and in this way be able to obtain the specific anisotropy shape for this model, which is necessary to introduce into the total radial force and establish the analysis.

\subsubsection{Both polytropes with $\gamma\ne 1$}
In this subsection we shall assume that $\gamma_r \ne 1$,  $\gamma_\bot\ne 1$, and the tangential pressure satisfies the polytropic equation of state 
\begin{eqnarray}\label{Pt}
P_{\perp}=K_{\perp}\rho^{\gamma_{\perp}},
\end{eqnarray}
whereas the radial pressure satisfies (\ref{p2}).
Then, from (\ref{delta}) we have
\begin{eqnarray}\label{anis}
\Delta=K_{\perp}\rho^{\gamma_{\perp}}-K_{r}\rho^{\gamma_{r}}.
\end{eqnarray}
Introducing $\Psi$, $\alpha$ and the same definitions and conventions we we arrive to 
\begin{eqnarray}\label{deltap}
\Delta=P_{rc}(\Psi^{n_{r}\gamma_{\perp}}-\Psi^{n_{r}\gamma_{r}})=\rho_{c}\alpha \Psi^{n_{r}}(\Psi^{\theta}-\Psi),
\end{eqnarray}
where $\theta=n_{r}/n_{\perp}$. Thus Eq. (\ref{trf}) leads to 
\begin{eqnarray}
\mathcal{\hat{R}} &=& \Bigg\{ \Psi^{n_r} \frac{d\Psi}{d\xi}+\frac{\alpha \Psi^{n_r+1}}{a(\xi)}\Bigg[\frac{c(\xi)}{\xi^2}\Bigg(1+\frac{n_r}{\alpha(n_r+1)\Psi}\Bigg) \nonumber \\ &+& \xi \Psi^{n_r}b(\xi)\Bigg] + \frac{\alpha n_r\Psi^{n_r}\eta}{\xi^2(n_r+1)}\frac{b(\xi)d(\xi)}{a(\xi)^2}\Bigg\} \delta\phi \nonumber \\ &-& \frac{2}{(n_r+1) \xi}  \Psi^{n_{r}}(\Psi^{\theta}-\Psi) \delta \beta. 
\end{eqnarray}

\begin{figure}[ht!]
\centering
\includegraphics[scale=0.5]{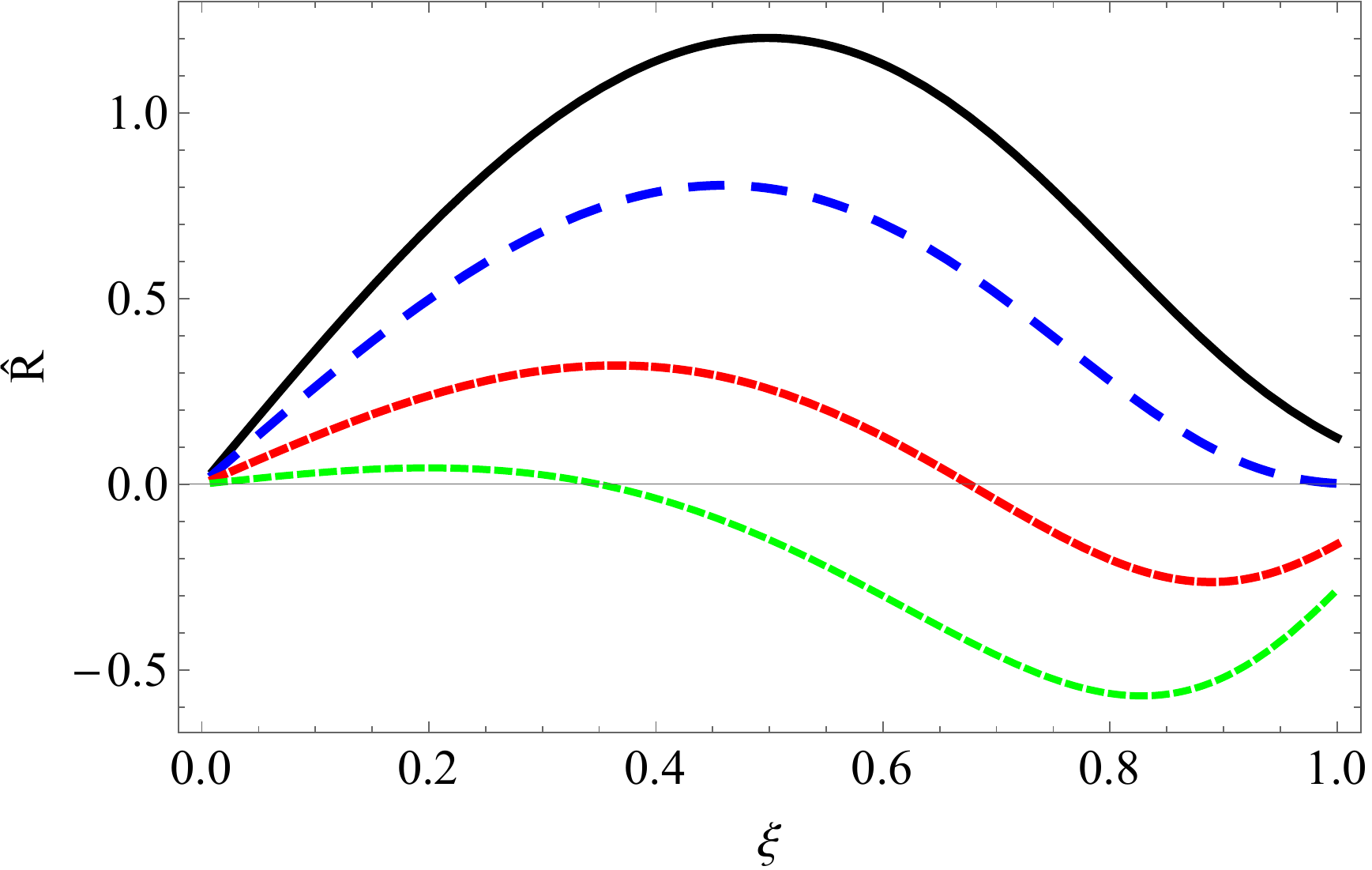}
\caption{\label{c-dp-dif-n-gamma-var}
$\mathcal{\hat{R}}$ as function of $\xi/\xi_{\Sigma}$, for 
$n_{r}=0.5$, $\alpha=1$, $\theta=0.5$ and $\Gamma=-2.5$ (black line), $\Gamma=-1$ (blue line), $\Gamma=1$ (red line) and $\Gamma=2.5$ (green line)
}
\end{figure}

\begin{figure*}[ht!]
\centering
\includegraphics[width=0.4\textwidth]{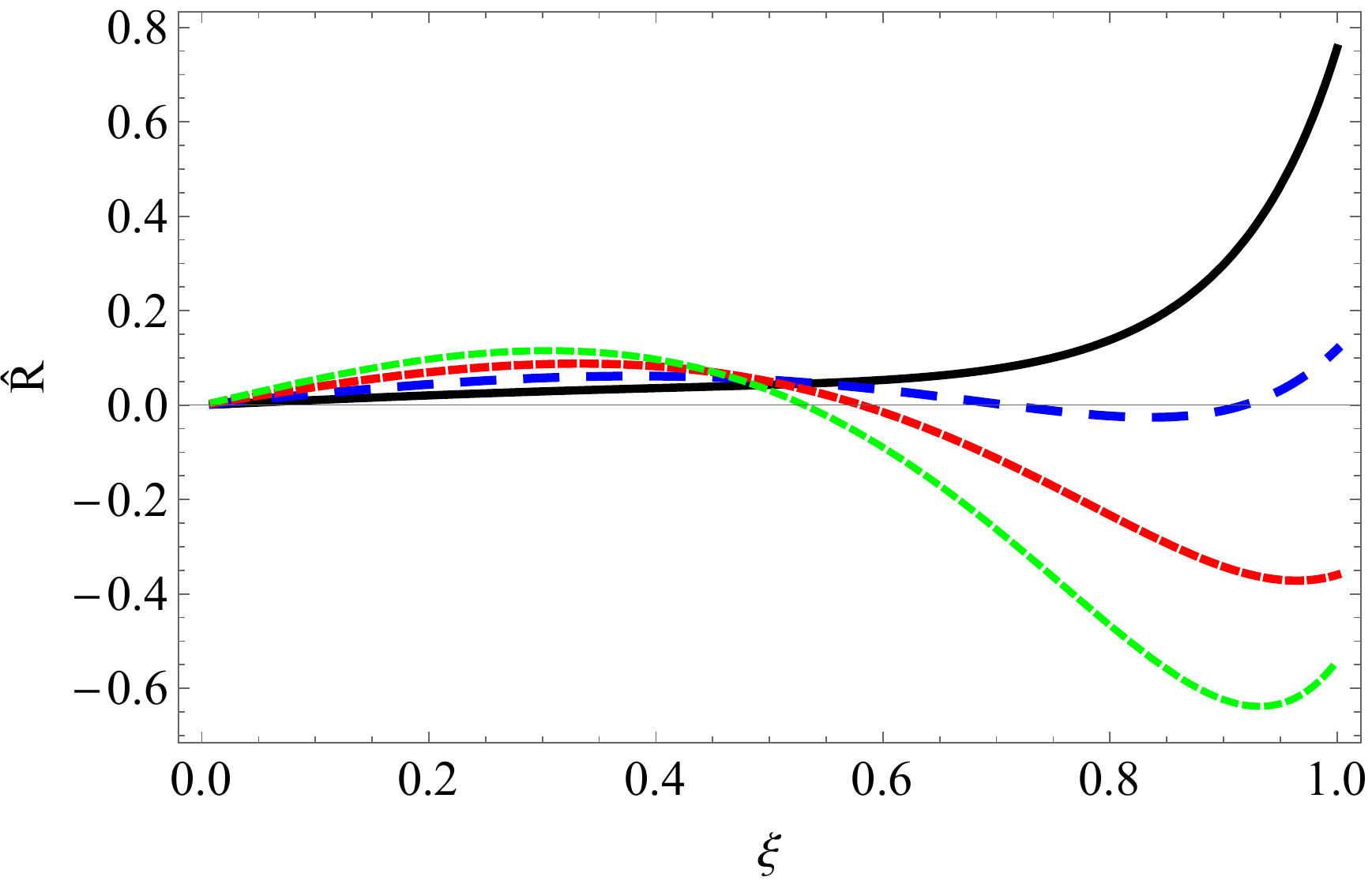}  \
\includegraphics[width=0.4\textwidth]{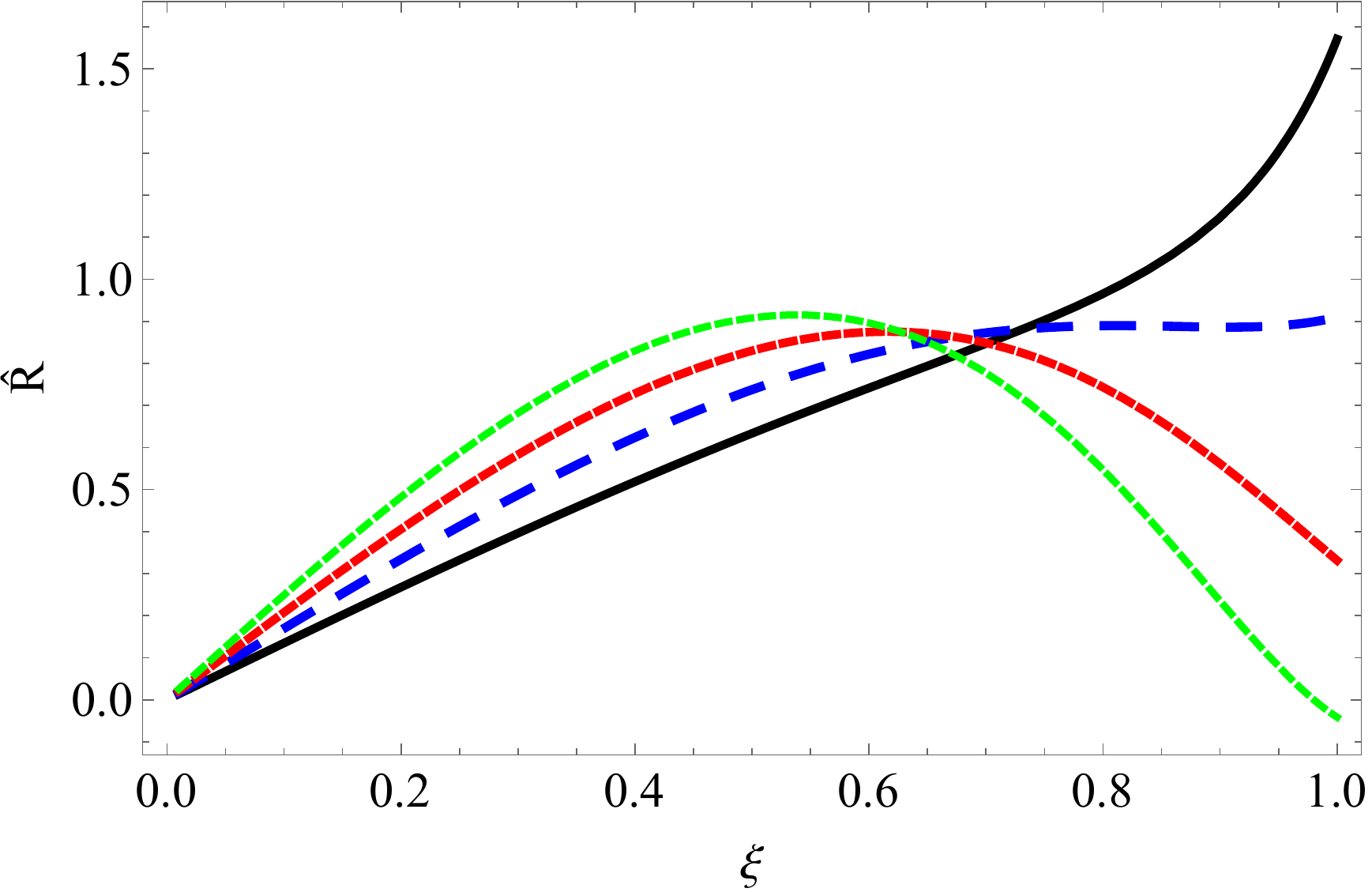}  \
\caption{
\label{c-dp-dif-n-alfa-var}$\mathcal{\hat{R}}$ as function of $\xi/\xi_{\Sigma}$, for $n_{r}=0.3$, $\theta=0.5$ and $\alpha=0.6$ (black line), $\alpha=0.7$ (blue line), $\alpha=0.8$ (red line) and $\alpha=0.9$ (green line)} with $\Gamma=1.5$ (left panel) and $\Gamma=-1.5$ (right panel)
\end{figure*}

\begin{figure*}[ht!]
\centering
\includegraphics[width=0.4\textwidth]{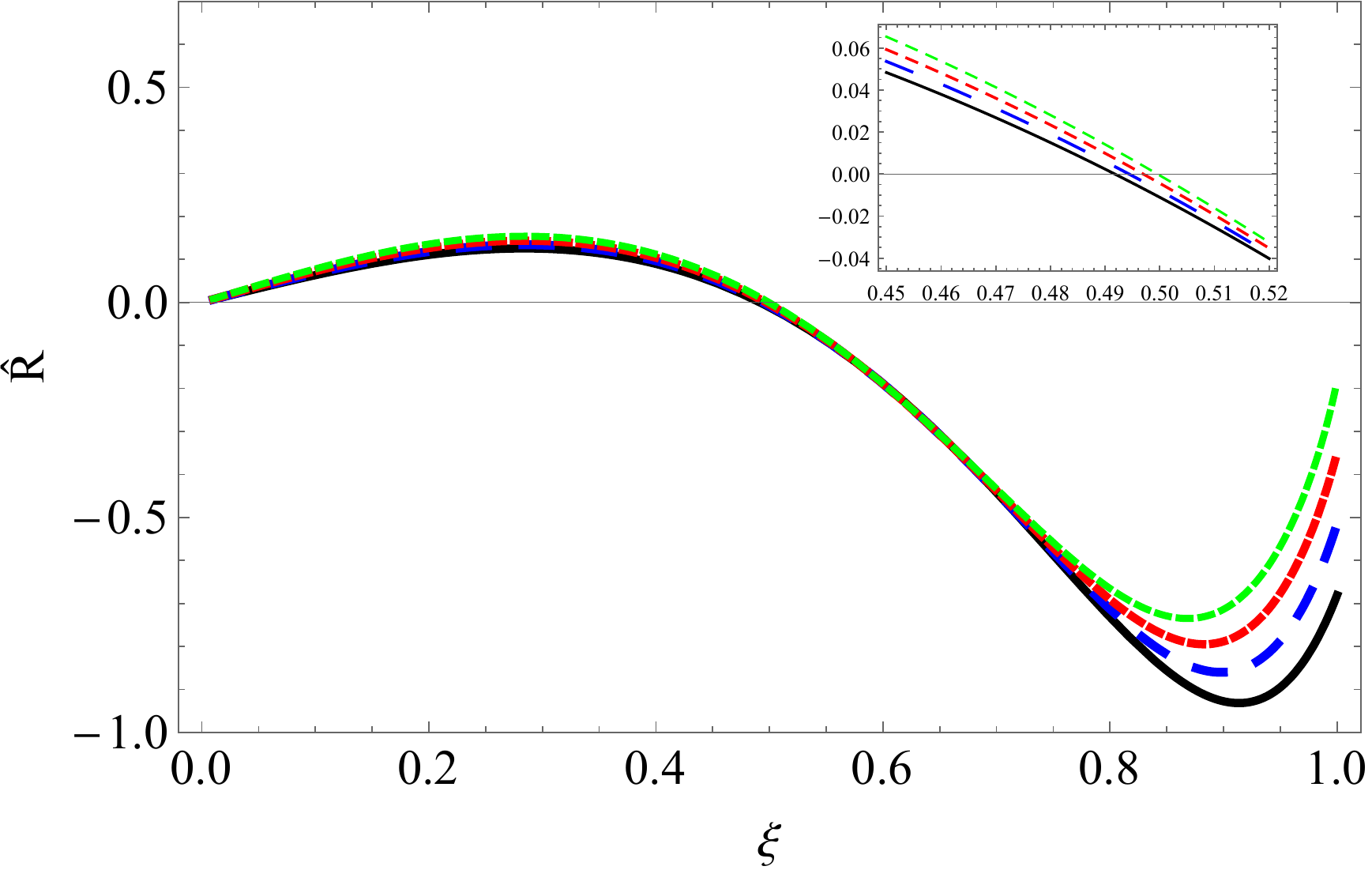}  \
\includegraphics[width=0.4\textwidth]{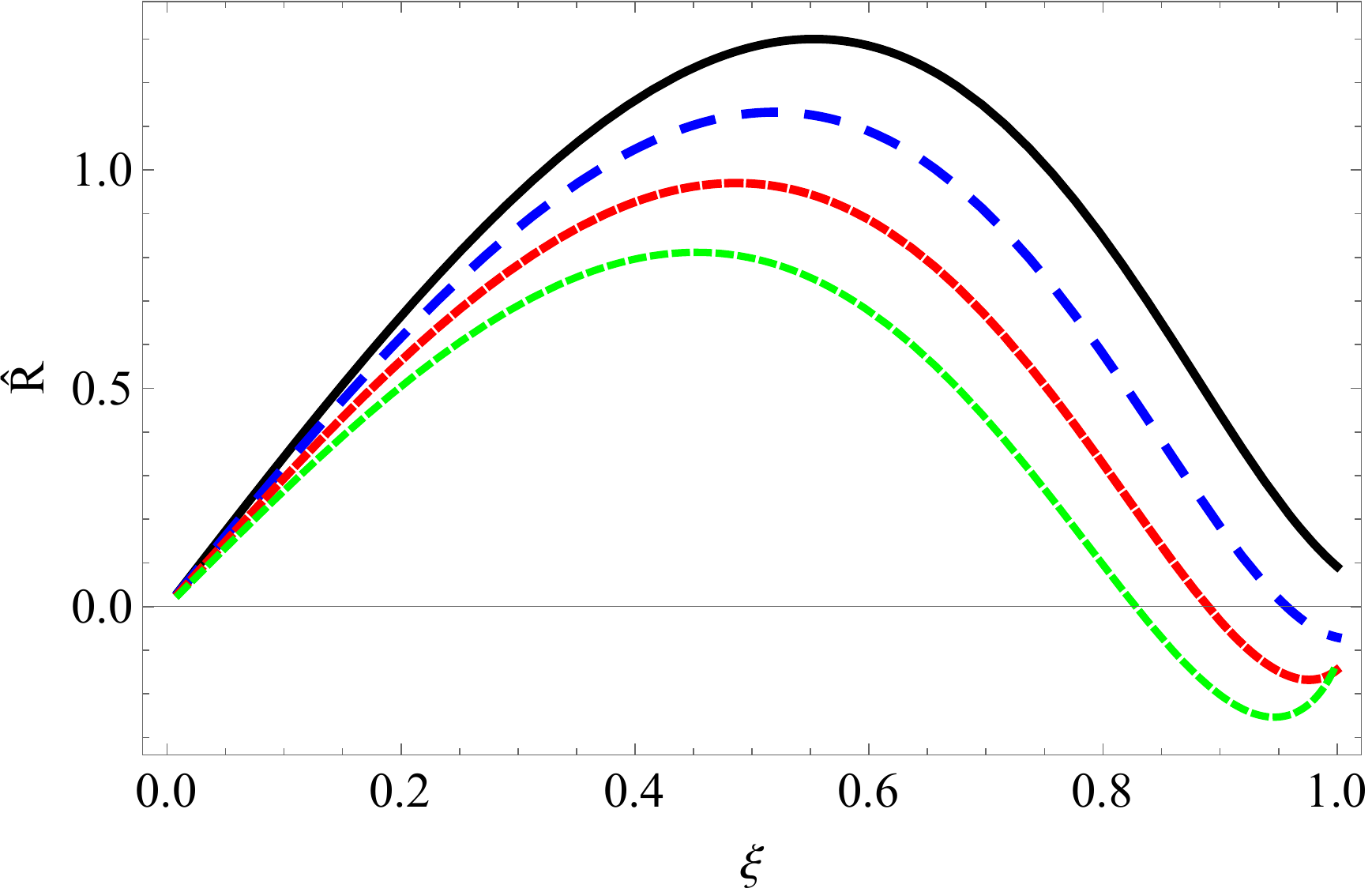}  \
\caption{
\label{c-dp-dif-n-theta-var}$\mathcal{\hat{R}}$ as function of $\xi/\xi_{\Sigma}$, for $n_{r}=0.3$, $\alpha=1$, $\theta=0.3$ (black line), $\theta=0.4$ (blue line), $\theta=0.5$ (red line) and $\theta=0.6$ (green line)} with $\Gamma=1.5$ (left panel) and $\Gamma=-1.5$ (right panel)
\end{figure*}

\begin{figure}[ht!]
\centering
\includegraphics[scale=0.5]{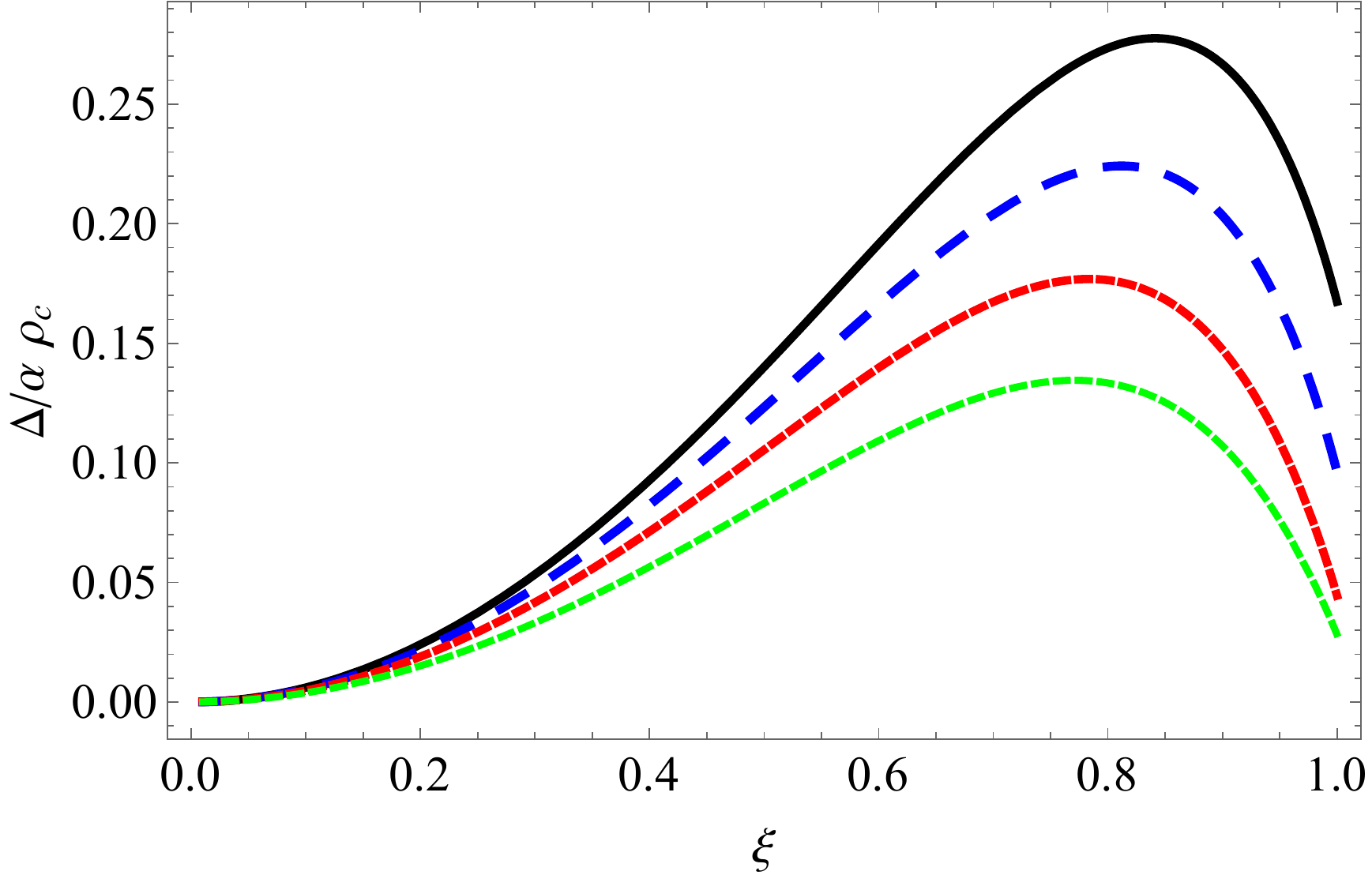}
\caption{\label{a-dp-n-dif-n-var}
Anisotropy $\Delta/\alpha\rho_{c}$ as function of $\xi/\xi_{\Sigma}$, for $\alpha=1$, $n_{r}=0.3$ and $\theta=0.3$ (black line), $\theta=0.4$ (blue line), $\theta=0.5$ (red line) and $\theta=0.6$ (green line)
}
\end{figure}

In Figs. \ref{c-dp-dif-n-gamma-var}, \ref{c-dp-dif-n-alfa-var}, \ref{c-dp-dif-n-theta-var}
it is shown the behaviour of $\mathcal{\hat{R}}$ for different values of the parameters in the legend. Also, to complement the discussion we plot in Fig. \ref{a-dp-n-dif-n-var} the behaviour of the anisotropy as a function of $\theta$. We see from expression (\ref{deltap}) that $\theta=n_{r}/n_{\perp}$ constitutes a parameter that allows us to control the anisotropy of the system in the case  where both polytropes have $\gamma\ne 1$. In reference \cite{6p} it was shown that if $\theta=1$ the system is isotropic. 

 
\subsubsection{Radial polytrope with $\gamma_r\ne 1$ and tangential polytrope with $\gamma_\bot= 1$}
In this case we assume
$P_{r}=K_{r}\rho^{1+\frac{1}{n_{r}}}$ with $\rho=\rho_{c}\Psi^{n_{r}}$ and
\begin{eqnarray}
P_{\perp}=K_{\perp}\rho,
\end{eqnarray}
from where the anisotropy reads
\begin{eqnarray}\label{a22}
\Delta=P_{rc}\Psi^{n_{r}}(1-\Psi).
\end{eqnarray}
So, we have obtained the specific form of the anisotropy for this case of the double polytrope model \cite{6p}. This will allow us to complete the perturbative scheme exposed before and then study the possibility of cracking when the fluid configuration is taken out of hydrostatic equilibrium. Now, the total radial force after the perturbation is
\begin{eqnarray}
\mathcal{\hat{R}} &=& \Bigg\{ \Psi^{n_r} \frac{d\Psi}{d\xi}+\frac{\alpha \Psi^{n_r+1}}{a(\xi)}\Bigg[\frac{c(\xi)}{\xi^2}\Bigg(1+\frac{n_r}{\alpha(n_r+1)\Psi}\Bigg) \nonumber \\ &+& \xi \Psi^{n_r}b(\xi)\Bigg] + \frac{\alpha n_r\Psi^{n_r}\eta}{\xi^2(n_r+1)}\frac{b(\xi)d(\xi)}{a(\xi)^2}\Bigg\} \delta\phi \nonumber \\ &-& \frac{2}{(n_r+1) \xi}  \Psi^{n_{r}}(1-\Psi) \delta \beta. 
\end{eqnarray}

\begin{figure}[ht!]
\centering
\includegraphics[scale=0.5]{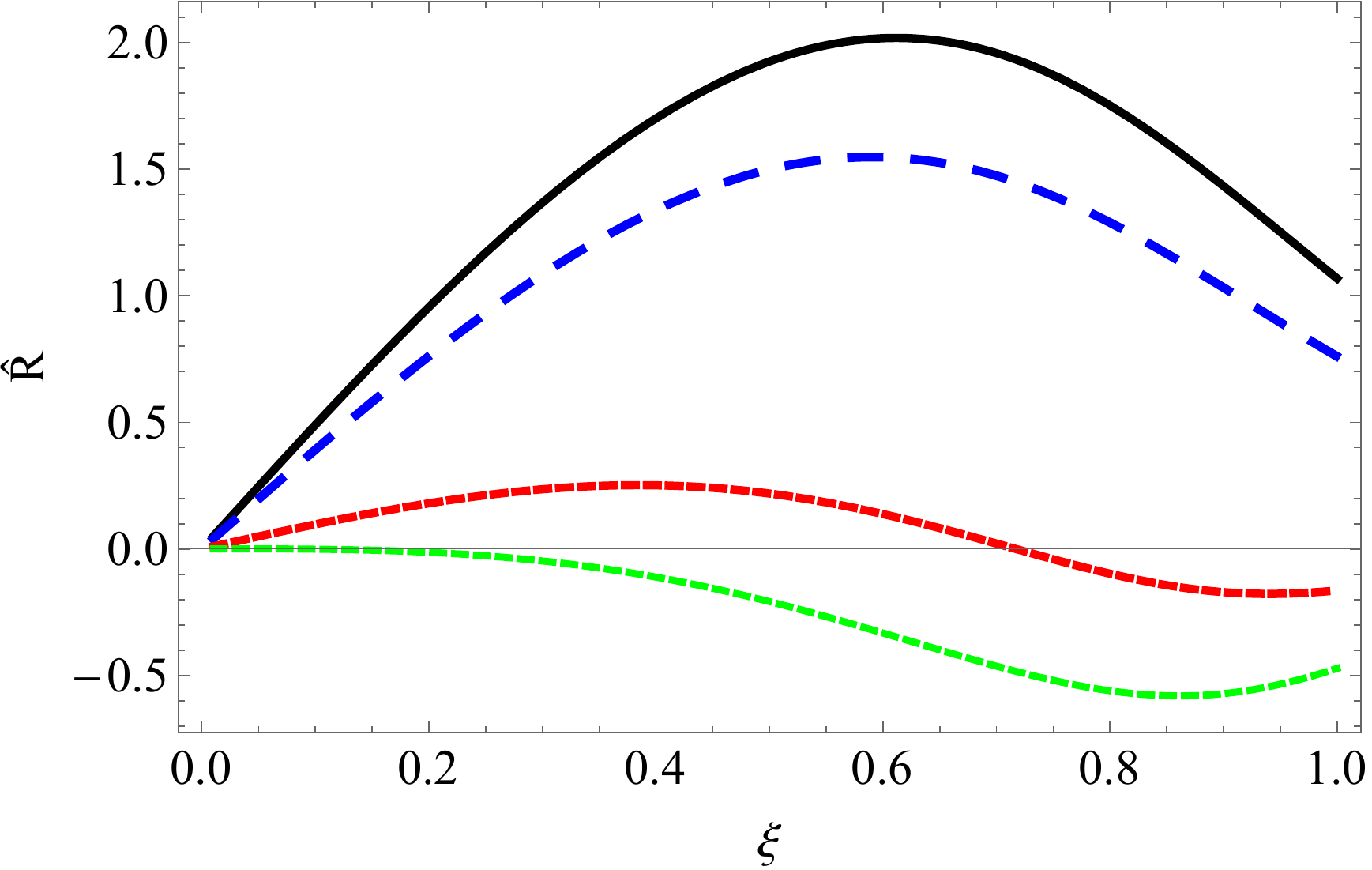}
\caption{\label{c-dp-gamma-var}
$\mathcal{\hat{R}}$ as function of $\xi/\xi_{\Sigma}$, for 
$n_r =0.7$, $\alpha=1$,  and $\Gamma=-2.5$ (black line), $\Gamma=-1.5$ (blue line), $\Gamma=1.5$ (red line) and $\Gamma=2.5$ (green line)
}
\end{figure}

\begin{figure*}[ht!]
\centering
\includegraphics[width=0.4\textwidth]{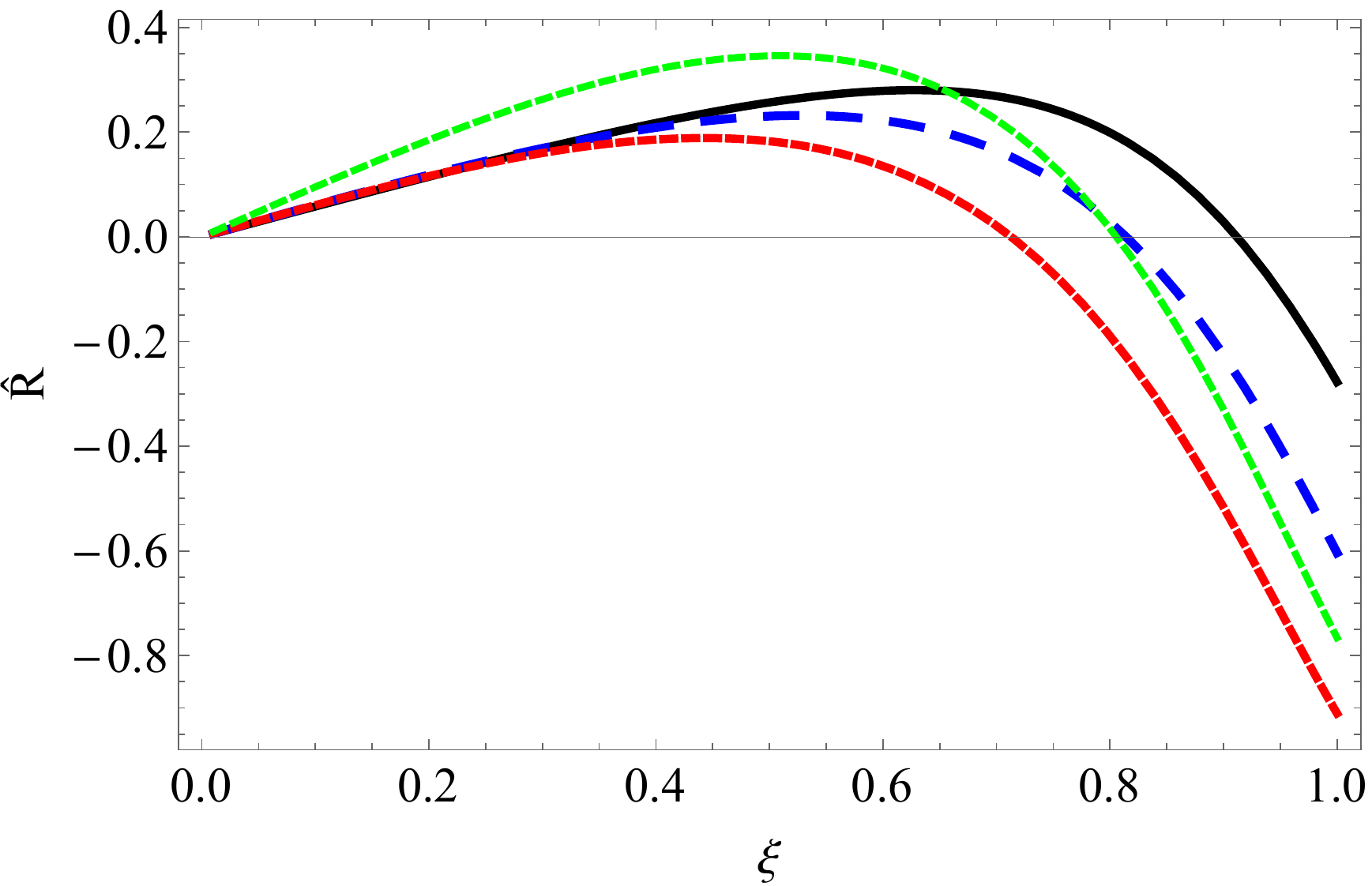}  \
\includegraphics[width=0.4\textwidth]{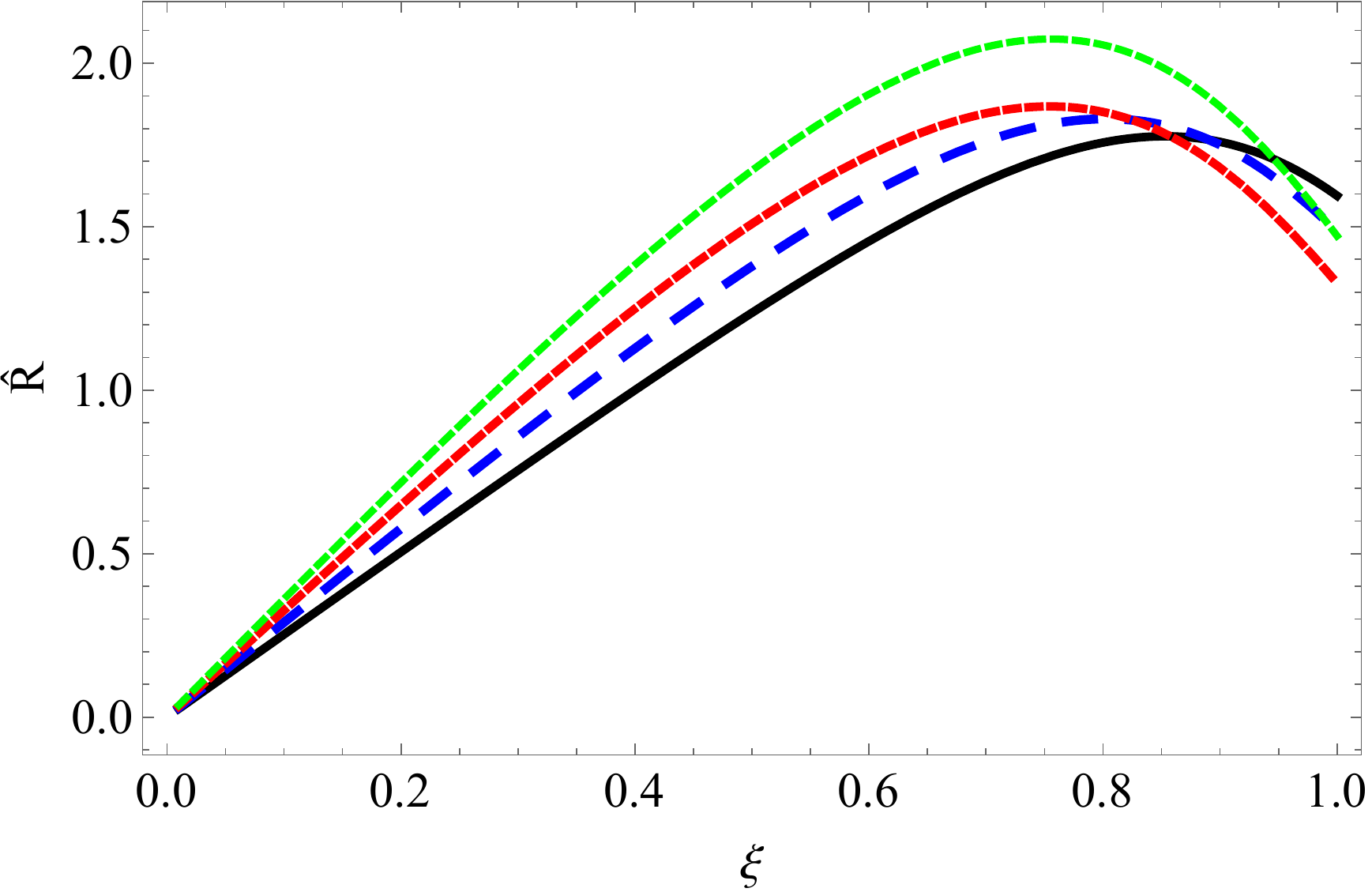}  \
\caption{\label{c-dp-alfa-var}
$\mathcal{\hat{R}}$ as function of $\xi/\xi_{\Sigma}$, for 
$n_r =0.3$, and $\alpha=0.6$ (black line), $\alpha=0.7$ (blue line), $\alpha=0.8$ (red line) and $\alpha=0.9$ (green line) with $\Gamma=1.5$ (left panel) and $\Gamma=-1.5$ (right panel).
}
\end{figure*}

\begin{figure*}[ht!]
\centering
\includegraphics[width=0.4\textwidth]{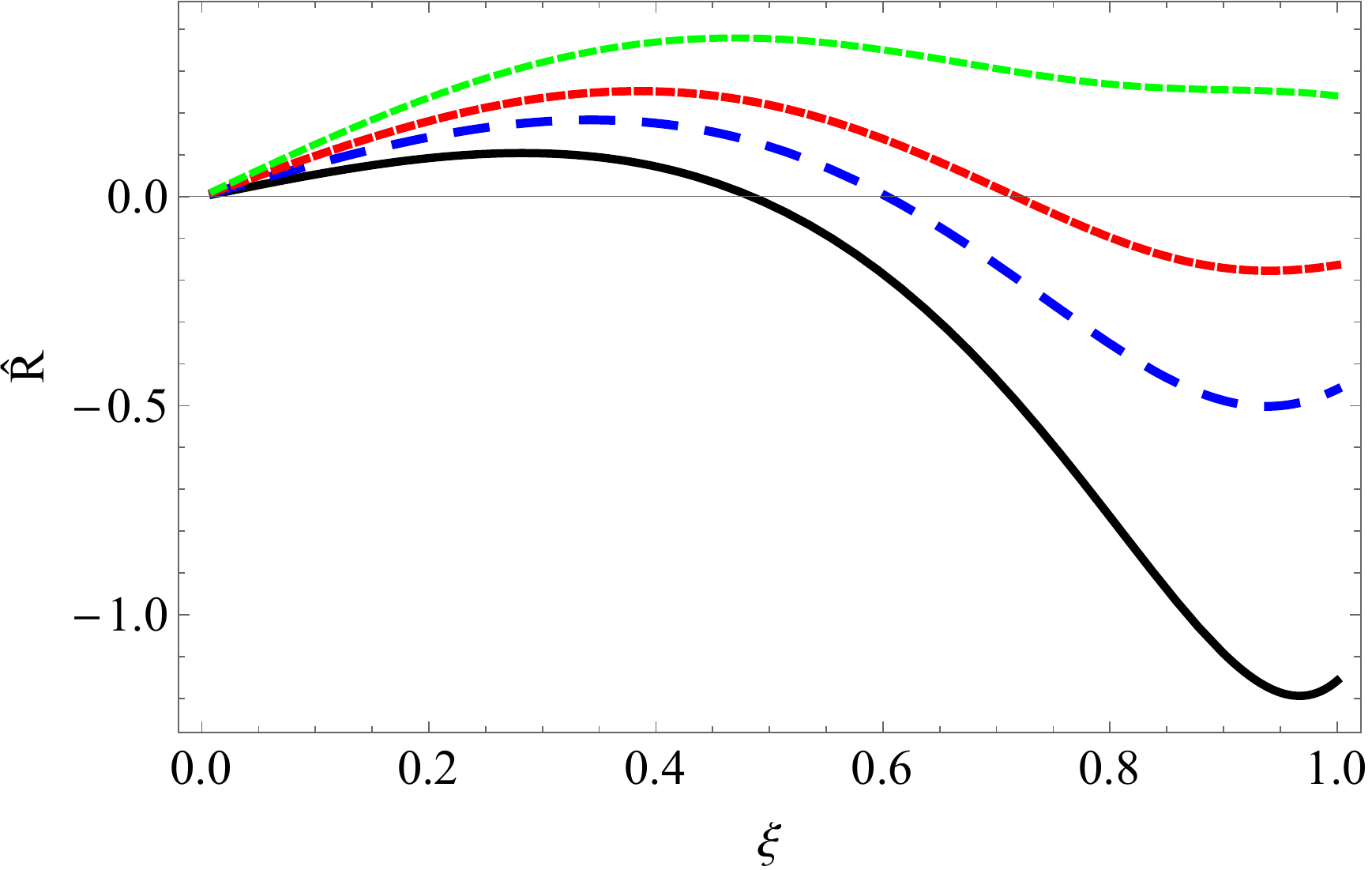}  \
\includegraphics[width=0.4\textwidth]{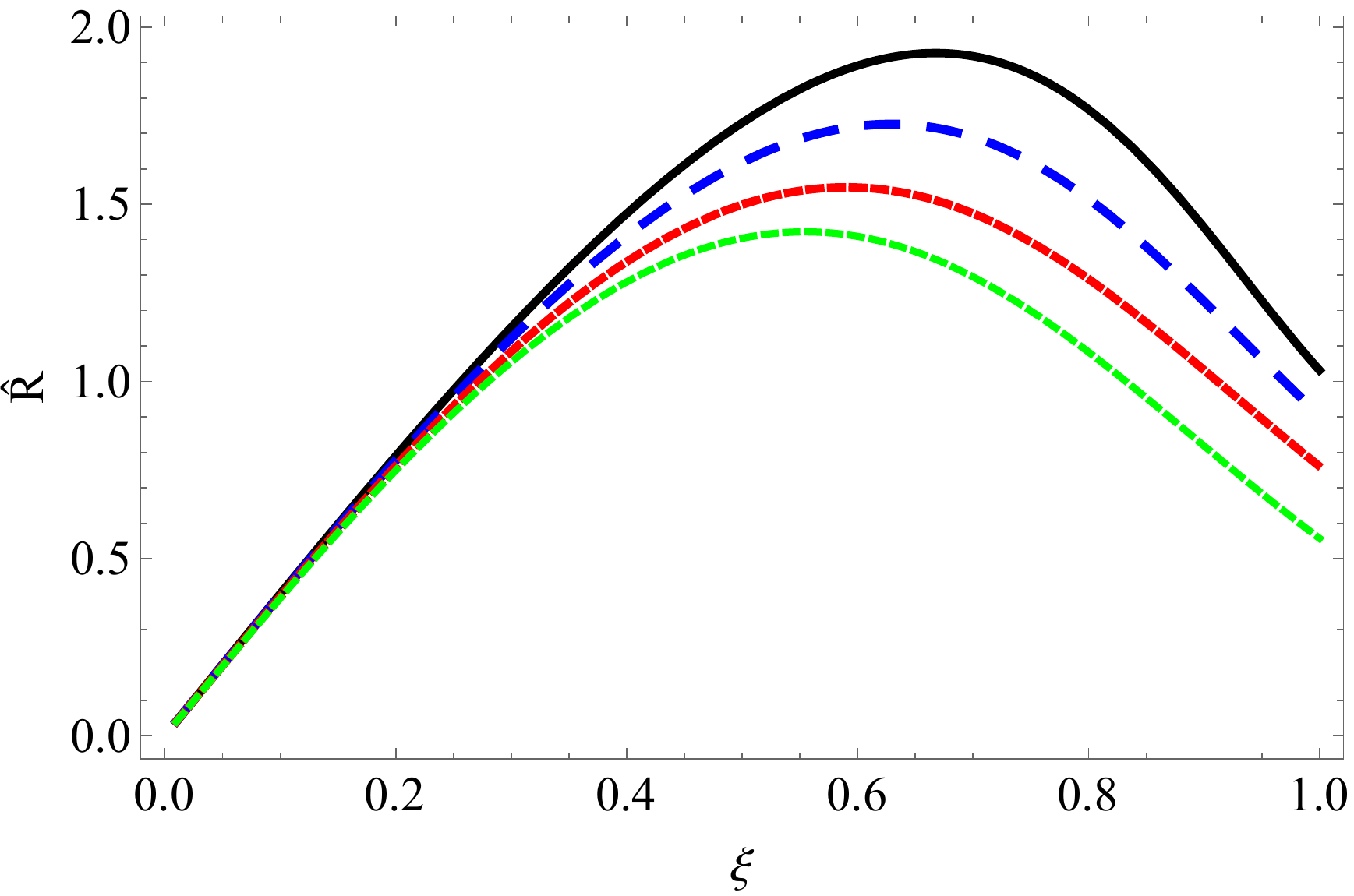}  \
\caption{\label{c-dp-n-var}$\mathcal{\hat{R}}$ as function of $\xi/\xi_{\Sigma}$, for 
 $\alpha=1$,  and $n_r =0.3$ (black line), $n_r =0.5$ (blue line), $n_r =0.7$ (red line) and $n_r =0.9$ (green line) with $\Gamma=1.5$ (left panel) and $\Gamma=-1.5$ (right panel)
}
\end{figure*}

\begin{figure}[ht!]
\centering
\includegraphics[scale=0.5]{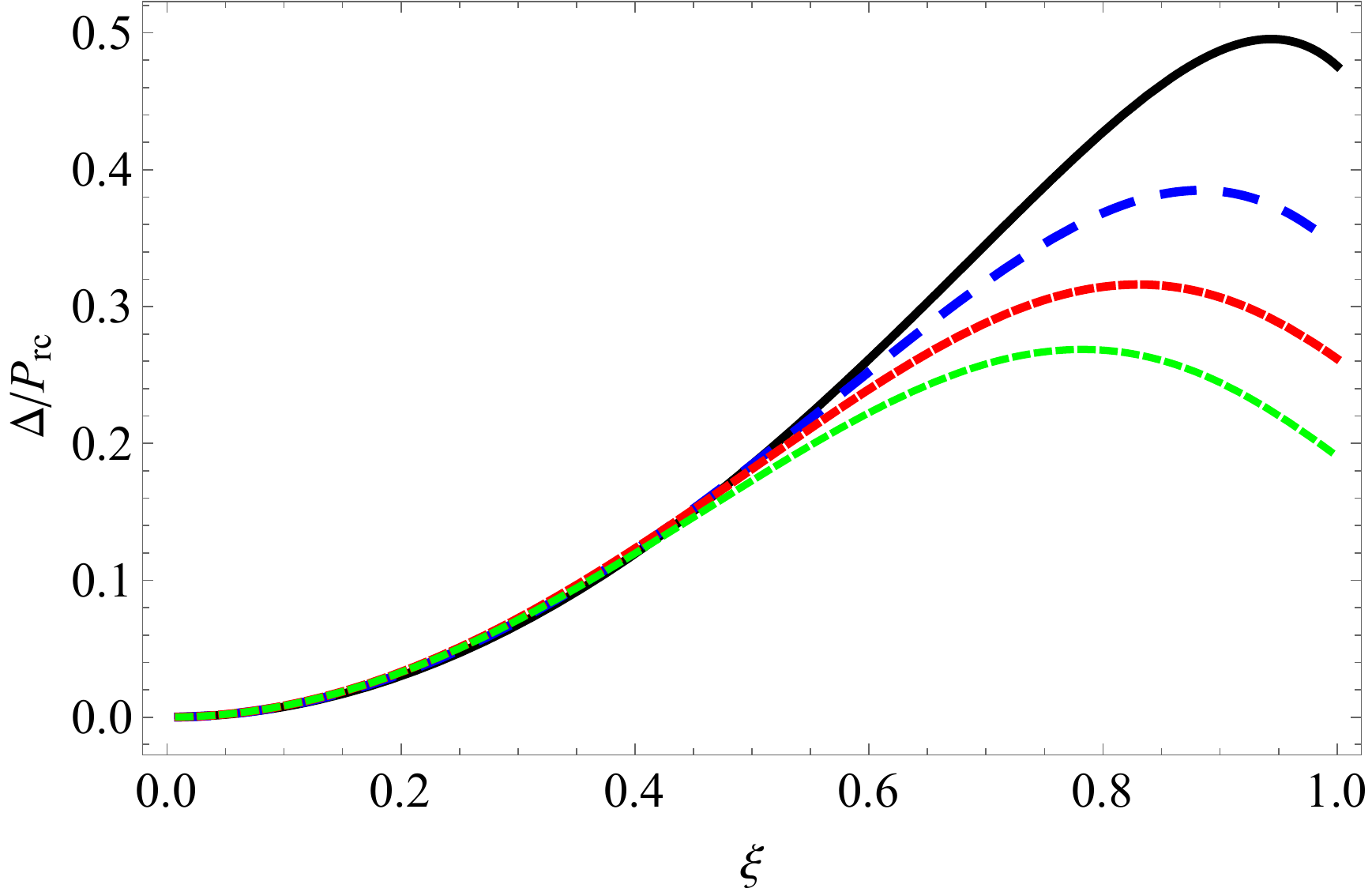}
\caption{\label{a-dp-n-var}
Anisotropy $\Delta/\alpha\rho_{c}$ as function of $\xi/\xi_{\Sigma}$, for $\alpha=1$ and $n_r =0.3$ (black line), $n_r =0.5$ (blue line), $n_r =0.7$ (red line) and $n_r=0.9$ (green line)}
\end{figure}

In Figs. \ref{c-dp-gamma-var}, \ref{c-dp-alfa-var}, \ref{c-dp-n-var}
we show the behaviour of $\mathcal{\hat{R}}$ for different values of the parameters. To complement the discussion we plot the behaviour of the anisotropy (\ref{a22}) as a function of the polytropic index $n_{r}$, shown in Fig. \ref{a-dp-n-var}. Note that in this case (radial polytrope with $\gamma_r\ne 1$ and tangential polytrope with $\gamma_\bot= 1$) the anisotropy is controlled only by the radial polytropic index $n_r$. 

\subsection{The vanishing complexity polytrope}

A new definition of complexity for spherically symmetric static self-gravitating fluids, in the context of general relativity, was introduced in \cite{VC}. The scalar function, $Y_{TF}$, that arises from the interpretation of the orthogonal splitting of the Riemann tensor, describes how the local anisotropy of pressure and density inhomogeneity modify the value of the Tolman mass, with respect to its value for the homogeneous isotropic fluid, so it constitutes a good quantity to be defined as a complexity factor. A simple calculation \cite{VC,LH-C3} allows us to express $Y_{TF}$ in terms of the inhomogeneity of the energy density and the local anisotropy of the system like, 
\begin{eqnarray} \label{YTF2}
Y_{TF} = - 8\pi \Delta - \frac{4\pi}{r^3}\int^{r}_{0} \tilde{r}^3 \rho' d\tilde{r}.
\end{eqnarray}
According to (\ref{YTF2}), the vanishing complexity factor condition, reads:
\begin{eqnarray} \label{YTF=0}
Y_{TF}=0\quad \Longrightarrow \quad \Delta = - \frac{1}{2 r^3}\int^{r}_{0} \tilde{r}^3 \rho' d\tilde{r},
\end{eqnarray}
that may be regarded as a non–local equation of state. Now, integrating this expression by parts and using the definition of the mass function, it is a  straightforward computation to show 
\begin{equation}
    \Delta = \frac{1}{8\pi r^3} \left[3m-r\frac{dm}{dr}\right],
\end{equation}
which can be written as 
\begin{eqnarray}
\Delta = \frac{r}{16\pi}\left(\frac{e^{-\lambda}-1}{r^2}\right)',
\end{eqnarray}
that, with the exception of a $1/2$ factor, constitutes the same expression for the anisotropy obtained for the conformally flat polytrope. Therefore, this anisotropy function has essentially the same behavior as that shown in Fig. \ref{a-cf-n-var}. Thus, when we adopt the polytropic equation of state, the total radial force becomes 
\begin{eqnarray}
\mathcal{\hat{R}} &=& \Bigg\{ \Psi^{n} \frac{d\Psi}{d\xi}+\frac{\alpha \Psi^{n+1}}{a(\xi)}\Bigg[\frac{c(\xi)}{\xi^2}\Bigg(1+\frac{n}{\alpha(n+1)\Psi}\Bigg) \nonumber \\ &+& \xi \Psi^{n}b(\xi)\Bigg] + \frac{\alpha n\Psi^{n}\eta}{\xi^2(n+1)}\frac{b(\xi)d(\xi)}{a(\xi)^2}\Bigg\}\; \delta\phi \nonumber \\ &-& \frac{1}{\alpha (n+1) \xi}  \left(\frac{3\eta}{\xi^3}-\Psi^n\right)\; \delta \beta,
\end{eqnarray}
which in essence has the same behavior as the case exposed for the conformally flat polytrope so we will not repeat the analysis.\\

\section{Results}


Figures \ref{c-cf-gamma-var}, \ref{c-dp-dif-n-gamma-var} and \ref{c-dp-gamma-var} show the dependency of the total radial force for different values of the $\Gamma$ which according to (\ref{Gamma}) provides the relationship between the parameters $\phi$ and $\beta$ used for perturbing the physical system.  We observed that when $\Gamma$ increases the radius where the total radial force has a change of sign
(cracking surface) moves towards deeper regions inside the compact object. This is a recurrent fact for all our models. Besides, in some cases,
we may have absence of a cracking for
negative value of $\Gamma$ which indicates that the system is stable in this sense. For example, this occur for $\Gamma=-5$ (black line in figure \ref{c-cf-gamma-var}) for the conformally flat polytrope model, $\Gamma=-2.5$ (black line in figure \ref{c-dp-dif-n-gamma-var}) for the case 1-double polytrope, and $\Gamma=-2.5$, $\Gamma=-1.5$ (black and blue lines in figure \ref{c-dp-gamma-var}) for the case 2-double polytrope model.\\

A similar behavior is obtained, for all the models, when the dependence of $\mathcal{\hat{R}}$ with respect to $\alpha=\frac{P_{rc}}{\rho_{c}}$ is studied. The situation described in figures \ref{c-cf-alfa-var}, \ref{c-k-alfa-var}, \ref{c-dp-dif-n-alfa-var} and \ref{c-dp-alfa-var} is representative for a wide range of parameters (for which there exist bounded configurations satisfying the required physical conditions). Note that the cracking radius ($\mathcal{\hat{R}}=0$) moves to deeper regions of the object as $\alpha$ increases. Furthermore, we observe that in some cases for small values of $\alpha$ there is no cracking: see the black line ($\alpha=0.88$) and blue line ($\alpha=0.90$) in figure \ref{c-k-alfa-var} for the Karmarkar model. Also, the black line ($\alpha=0.60$) in figure \ref{c-dp-dif-n-alfa-var} for the case 1-double polytrope model. Then, for smaller $\alpha$, the critical radius moves towards the surface of the object until it does not produce cracking at all. It seems that the strongest and deepest crackings are associated with the largest values of $\alpha$. The right panel of figures \ref{c-cf-alfa-var}, \ref{c-dp-dif-n-alfa-var} and \ref{c-dp-alfa-var} show that for negative values of $\Gamma$ there is no cracking (for all $\alpha$), indicating stability against this phenomena for that relationship of the parameters involved.\\

In figures \ref{c-cf-n-var}, \ref{c-k-n-var} and \ref{c-dp-n-var} it is exhibited the relationship existing between the occurrence of cracking and the polytropic index $n$. For the conformally flat model, the critical radius moves to more internal regions as $n$ increases, (see figure \ref{c-cf-n-var}) showing deeper and softer cracking for these values. Instead, in the case of the Karmarkar polytrope, figure \ref{c-k-n-var}, and the case 2-double polytrope, figure \ref{c-dp-n-var}, the behavior is totally the opposite. An increase in the polytropic index corresponds to a displacement of the cracking radius towards external regions of the compact object, where it is also observed that cracking is smoother. Even, we can have configurations that do not present cracking for a certain range of parameters and certain values of $n$. This is observed, for example, in the Karmarkar model, figure \ref{c-k-n-var} (green line), corresponding to $n=0.2$ and for the case 2-double polytrope model, figure \ref{c-dp-n-var} (green line), corresponding to $n=0.9$. Again, we observe that the system shows stability against cracking, for some negative $\Gamma$ values, in all cases (observe the right panel of the figures).\\

Finally, we have presented figures \ref{a-cf-n-var}, \ref{a-dp-n-dif-n-var} and \ref{a-dp-n-var}, in order to exhibit further the close relationship existing between the occurrence of cracking and the type of anisotropy displayed for each model. Figure \ref{a-cf-n-var} represents the anisotropy associate to the conformally flat model exposed for different values of the  polytropic index $n$. In figure \ref{a-dp-n-dif-n-var} we plot the behaviour of the anisotropy, for the case 1-double polytrope, as a function of $\theta$ and figure \ref{a-dp-n-var} shows the case 2-double polytrope, where the anisotropy is expressed by means of the polytrope index $n_r$. In all cases, an anisotropy appears as a properly increasing function.

We see in figure \ref{a-cf-n-var} that for greater values of $n$ the anisotropy grows. Comparing with figure \ref{c-cf-n-var} it seems to indicate that for greater anisotropy the cracking is more internal and less abrupt for the conformally flat model. In the same sense, figure \ref{a-dp-n-dif-n-var} shows the dependence of the local anisotropic function with respect to $\theta$ for the case 1-double polytrope model showing that an increase in $\theta$ implies a decrease of the anisotropy (mainly in the outer layers of the object). Relating this fact to the behavior observed in figure \ref{c-dp-dif-n-theta-var}, we get that external and stronger cracking are produced for smaller values of anisotropy and positive values of $\Gamma$ (left panel). The opposite occurs for negative values of $\Gamma$ (right panel): deeper and smoothed cracking are produced for smaller values of anisotropy. In figure \ref{a-dp-n-var} its observed that the anisotropy decreases with the increase value of $n_r$, mostly in the outer regions of the object, for the case 2-double polytrope model. In essence this behavior is similar to case 1-double polytrope model. Now, corresponding to the same external zone, we observe cracking shift towards outer regions (besides, cracking getting softer) of the stellar object, according to figure \ref{c-dp-n-var}. The right panel (negative $\Gamma$) of this figure shows again, the interesting result, of stability of the system against cracking.\\

It is important to mention that no graphical representation or analysis for the polytrope model with vanishing complexity was performed. The main reason is that the inherited anisotropy produced by the vanishing condition of the complexity parameter is identical (except for a global constant factor) to that of the conformally flat polytrope model. In fact, the total radial force $\mathcal{\hat{R}}$ is essentially the same to the corresponding first model of this work so qualitatively we have the same results. This, in itself, represents an interesting and non-trivial result worth reporting.

\section{Conclusions}

It is important to stress that the occurrence of cracking, has direct implications on the structure and evolution of the compact object, only at time scales that are smaller than, or equal to, the hydrostatic time scale. This is so because, as already mentioned, what we do is to take a “snapshot” just after the system leaves the equilibrium. To find out whether or not the system will return to the state of equilibrium afterward would require an integration of the evolution equations in the dynamic case for a finite period of time (greater than hydrostatic time). Nevertheless, it is clear that the occurrence of cracking would drastically affect the future structure and evolution of the compact relativistic object.\\

We have investigated the conditions under which, general relativistic polytropes for anisotropic matter, exhibit cracking (and/or overturning), when submitted to fluctuations of energy density and anisotropy. To achieve this, a general and systematic method was proposed to study the departure from equilibrium for any internal, anisotropic, and spherically symmetric solution of Einstein field equations. This method has the advantage of being independent of the particular characteristics (parameters) of the considered models (see \cite{cn4} for previous developments). Thus, we have shown that cracking occur for a wide range of the parameters. For various types of polytropes, the main conclusions are basically the same, namely: the strongest and deepest crackings occur for bigger values of the parameters $\Gamma$ and $\alpha$. Also, most of the polytrope models seek to stabilize when $\Gamma$ grows negatively. This result is also observed for small values of $\alpha$, which represents an interesting fact (since the $\alpha$ parameter is related with the non-relativistic limit). The anisotropy of the pressure, depending of the case, can be ``controlled'' by the polytrope index $n$ or the $\theta$ parameter. A distinct feature of all the models studied here, is the fact that the ``core'' (the inner part), and the ``envelope'' (the outer part),  respond differently to different degrees of anisotropy. This fact was already pointed out in \cite{2p,4p,HRW}.\\

From the first moments in which cracking was defined \cite{cn, cnbis, cn1, cn2, cn4}, some hypotheses of possible applications to different types of astrophysical events were raised, and so there are several indications that allow us to assume such a fact. We are modeling an event that could take place in a very compact object where general relativity and the presence of local anisotropy plays a predominant role. Furthermore, it has been shown that cracking results only if, in the process of perturbation leading to departure from equilibrium, the local anisotropy is perturbed. The number of physical processes giving rise to small deviations from local isotropy and variations of local anisotropy in the high density regime is quite large. The first one is the intense magnetic field observed in compact objects such as white dwarfs and neutron stars. These, in turn can also cause fluctuations in the anisotropy of the system. It is well known that for various types of astrophysical objects, magnetic field lie at the heart of stellar dynamics and possess activity cycles which can involve a significant fraction of the energy budget of the star. However, attempts to relate the observed characteristics of late-type main sequence stars to a global description of their magnetic properties is complicated. Nevertheless, the are several reports that indicate that there is a very little room for the formation of nonmagnetic neutron stars from supernova events \cite{Lyne1, Lyne2, Srin} due to processes like dynamical field amplification in a carbon burning core \cite{Ruderman1}. Also, we have the possibility of neutrino viscosity produced by neutrino trapping or the existence of a solid core and the presence of type $p$ superfluid or type $II$ superconductivity can be invoked as possible sources of local anisotropy. There are also exotic phase transitions, that may occur during the process of gravitational collapse \cite{CoP}, like the transition to a pion condensate \cite{HSS}. Finally, the superposition of two perfect fluids may be formally described as an anisotropic fluid. This scheme allows one to evaluate the fractional anisotropy in a neutron star due to the contamination of electrons and protons required to stabilize neutron matter against $\beta$ decay \cite{Baylin}. \\

It is important to mention that another source of dynamic instability can occur in astrophysical scenarios due to the violation of Harrison-Zeldovich-Novikov stability condition (see for example \cite{Kokkotas}). This criterion implies that as the central density of a model increases, so does its total mass. As a consequence, this criterion finds a critical point that separates stable configurations from unstable ones when the total mass decreases with increasing central density. This is relevant, since the existence of the Tolman-Oppenheimer-Volkoff limit is known for more compact configurations than white dwarfs. The best predictions about local physical characteristics come from modeling these objects through a degenerate cold Fermi gas. Fermi temperature fluctuations can produce fluctuations in density, especially if we consider complicated nuclear reactions and even unknown internal processes can be of relevance.  \\

We would like to conclude this work by speculating about possible scenarios where the occurrence of cracking might be invoked, in order to understand the related observational data. One of these situations could be the collapse of a supermassive star. The occurrence of  cracking at  the inner core, would certainly change (in some cases probably enhance) the conditions for the ejection of the outer mantle in a supernova event. This will be so for both the ``prompt'' \cite{33, 34} and the ``long term'' mechanisms \cite{35, 36, 37, 38}.

Also, one is tempted to invoke cracking as the possible origin of quakes in neutron stars \cite{39, 40,  41}. In fact, large scale crust cracking in neutron stars and their relevance in the occurrence of glitches and bursts of x-rays and gamma rays have been considered in detail by Ruderman (see \cite{42} and references therein). Evidently, the characteristics of these quakes, and  those of the ensuing glitches, would strongly depend on the depth at which the cracking occurs. In this respect it is  worth noticing, the already mentioned fact, that the depth at which the cracking may appear, is highly dependent on the parameters $\alpha$ and $\Gamma$ and also on $n$ and $\theta$ (used to measure the anisotropy)  in our models. The specific facts are clearly dependent on each model.

However, neutron stars are rotating objects and in general the same causes that generate anisotropy (intense magnetic fields, Fermi fluid) may produce deviations from spherical symmetry, so the situation is clearly controversial. The assumption of relatively large scale crust cracking is a hypothesis which must still be supported mainly by comparisons of its consequences with neutron star observations. Anyway, we would like to emphasize that our aim here is not to model in detail any of the scenarios, but just call attention to the possible occurrence of cracking in such important configurations, as those satisfying a polytropic equation of state, and its relationship with fluctuations of local anisotropy. In this way, whatever the origin of the anisotropy would be (no matter how small) cracking may occur, and this fact would drastically affect the outcome of the evolution of the system.

\section{Acknowledgements}

P.L wants to say thanks for the financial support received by the Project ANT1956 of the Universidad de Antofagasta and CONICYT PFCHA / DOCTORADO BECAS CHILE/2019 - 21190517. P.L is grateful Semillero de Investigaci\'on SEM 18-02 from Universidad de Antofagasta and the Network NT8 of the ICTP.



\end{document}